\documentclass[a4paper]{aa}
\usepackage{graphicx}
\usepackage{amssymb}
\include{def}
\def\spose#1{\hbox to 0pt{#1\hss}}
\def\lta{\mathrel{\spose{\lower 3pt\hbox{$\mathchar"218$}}
     \raise 2.0pt\hbox{$\mathchar"13C$}}}
\def\gta{\mathrel{\spose{\lower 3pt\hbox{$\mathchar"218$}}
     \raise 2.0pt\hbox{$\mathchar"13E$}}}

\newcommand{\etal}{{et al.} }

\begin{document}


   \title{Spectral \& Photometric Evolution of Simple Stellar Populations at Various Metallicities}

   \author{J. Schulz \and U. Fritze - v. Alvensleben \and C.S. M\"oller \and K. J. Fricke}

   \institute{Universit\"atssternwarte G\"ottingen, Geismarlandstr. 11, 37083 G\"ottingen, 
Germany, http://www.uni-sw.gwdg.de/$\sim$galev/}
   \offprints{U. Fritze -- v. A., \email{ufritze@uni-sw.gwdg.de}}
   \mail{ufritze@uni-sw.gwdg.de}

   \authorrunning{J. Schulz \etal}
   \titlerunning{Spectral Evolution of SSPs}

\date{Received xxx November 2001 / Accepted xxx}

\abstract{
A new set of evolutionary synthesis spectra are presented for {\bf S}imple {\bf
S}tellar {\bf P}opulations ({\bf SSP}s) covering ranges in metallicity from $
0.02 \leq Z/Z_{\odot} \leq 2.5$ and ages from $4\cdot 10^6$ yr $\leq t
\leq 16$ Gyr. They are based on the most recent isochrones from the Padova
group that extend earlier models by the inclusion of the thermal pulsing AGB
phase for stars in the mass range $2 \, M_{\odot} \leq m \leq 7 \,
M_{\odot}$ in accordance with the fuel consumption theorem. We show that with
respect to earlier models, inclusion of the TP-AGB phase leads to significant
changes in the $(V-I)$ and $(V-K)$ colors of SSPs in the age
range from $10^8$ to $\gta 10^9$ yr. Using model atmosphere
spectra from Lejeune \etal (\cite{lej2}, \cite{lej}), we calculate the spectral
evolution of single burst populations of various metallicities covering the
wavelength range from 90 {\AA} ~through 160 $\mu$m. Isochrone spectra are
convolved with filter response functions to describe the time evolution of
luminosities and colors in Johnson, Thuan \& Gunn, Koo, HST, Washington and
Str\"omgren filters.\\ 
The models and their results are not only intended for use
in the interpretation of star clusters
but also for combination with any kind of dynamical galaxy formation and/or
evolution model that contains a star formation criterion. Moreover, the
evolution of these single burst single metallicity stellar populations is
readily folded with any kind of star formation -- and eventually chemical
enrichment -- history to describe the evolutionary spectral synthesis of
composite stellar populations like galaxies of any type with continuous or
discontinuous star formation.  For these latter purposes we also present the
time evolution of ejection rates for gas and metals for two different {\bf
I}nitial {\bf M}ass {\bf F}unctions ({\bf IMF}s) as well as cosmological and
evolutionary corrections for all the filters as a function of redshift for
$0 \leq z \leq 5$ and two different cosmologies.\\ 
Extensive data files are provided in the electronic version, at CDS, and at our above www-address. }

\maketitle


\section{Introduction}
\label{sec:Introduction}
SSPs are single burst single metallicity stellar populations like star clusters
with all their stars having the same age and the same metallicity. Evolutionary
synthesis models describe the evolution of these SSPs in terms of photometry and
spectra. They are of interest for several reasons. 1. They allow to test a set
of stellar evolutionary tracks against observations of star clusters from young
open clusters to old {\bf G}lobular {\bf C}lusters ({\bf GC}s). 2. They are used
for the interpretation of star clusters formed or forming in strong starbursts
like those accompanying galaxy interactions or mergers. Being formed from the
pre-enriched ISM of the merging galaxies, these young star clusters may cover a
wide range in metallicity. 3. Evolutionary models for individual stellar
generations can directly be folded -- in terms of luminosity or spectral
evolution -- with any {\bf S}tar {\bf F}ormation {\bf H}istory ({\bf SFH}) to
describe the composite stellar population of a galaxy. Or, the other way round,
a complete library of spectra for SSPs of various ages and metallicities
provides a theoretical basis for population synthesis models. Population
Synthesis models search for a particular linear combination of input spectra
that gives best agreement with the observed spectrum of a galaxy with unknown
SFH. 4. Evolutionary synthesis results for SSPs can be used in any kind of
dynamical galaxy formation and/or evolution code that involves a star formation
criterion. An example is presented in Contardo \etal (\cite{con}), where -- in
the context of a cosmological structure formation scenario -- an individual
galaxy is followed from the onset of {\bf S}tar {\bf F}ormation ({\bf SF}) in a
couple of subgalactic fragments to a state where it resembles the Milky Way in
many respects. Spatially resolved observable quantities are obtained from a
superposition of SSP spectrophotometric evolutionary model results.

A basic difficulty for single burst models as opposed to models for galaxies --
which always have extended and continuous SFHs -- is the discreteness of the
stellar mass spectrum for which evolutionary tracks have been
calculated. Lifetime differences between stars for which stellar evolutionary
tracks are available are of the order of $\geq 5 \cdot 10^8$ yr, i.e. much
longer than the age spread among the stars in SSP models with burst durations of
$\leq 10^5$ yr and also much longer than a typical timestep in the evolutionary
synthesis model. In late evolutionary phases, this causes large numbers of stars 
to collectively enter and leave distinct phases of
evolution with the effect of significant oscillations in colors and magnitudes. 
These oscillations in the photometric properties have been cured in different
ways. E.g., they can be smoothed {\sl a posteriori}. This has the advantage that
the intrinsic uncertainty of the method remains clearly visible. A smooth
spectral evolution, however, is difficult to obtain. 
A second possibility to start from a discrete set of stellar
tracks and obtain a reasonably smooth photometric evolution is the use of a
Monte Carlo method as described e.g. by Kurth \etal (\cite{kurth}).
This method works very well for a photometric description but requires a huge
number of Monte Carlo stars and timesteps to yield a smooth spectral
evolution. The third possibility is the use of isochrones obtained on the basis of
a grid of stellar evolutionary tracks by interpolating between equivalent
evolutionary stages. Once the definition of equivalent evolutionary stages is
taken for granted, this method provides a smooth evolution of SSPs, both in
terms of photometry and spectra.

Here we present a new set of evolutionary synthesis models for the photometric
and spectral evolution of SSPs of various metallicities $0.02 \leq Z/Z_{\odot} 
\leq 2.5$ over a Hubble time, based on the latest Padova isochrones (cf. Sect. 2.2). In this
sense, our models are complementary to Leitherer \etal's (\cite{leitherer99})
Starburst99 models which provide a very detailed description of the earliest
evolutionary phases ($\leq 1$ Gyr) of instantaneous burst populations ($=$
SSPs). In Sect. \ref{sec:Models-Input-Physics}, we present our models and the
input physics we use. Sect. \ref{sec:Spectr-Evol-SSPs} gives the spectral
evolution of SSPs for 5 different metallicities from $10^6$ yrs to $10^{10}$ yr
and a discussion of the effects of metallicity on the time evolution of the
spectra. In Sect. \ref{sec:Lumin-Evol-Colors}, we present the evolution of
photometric quantities for a number of widely used filter systems. In
Sect. \ref{sec:Theor-calibr-colors}, we compare our results to earlier models
and GC data. In Sect. \ref{sec:Cosm-Evol}, we present cosmological and
evolutionary corrections for our SSP models in 2 different cosmologies in a way
that can immediately be used by observers or included into dynamical galaxy
formation and evolutionary codes.

Only a few selected figures are presented here, the full tables for the time 
evolution of spectra, luminosities, colors, ejection rates, etc. of SSPs 
of all metallicities as well as the corresponding evolutionary and cosmological 
corrections as a function of redshift can be found at our webpage
(http://www.uni-sw.gwdg.de/$\sim$galev/).

Our isochrone synthesis code is able to describe the evolution of stellar
populations in galaxies with extended SFHs, as well. In this paper, however, we
only present SSP model results.

\section{Models and Input Physics}
\label{sec:Models-Input-Physics}
\subsection{Input physics: IMF}
Results in this paper are presented for two different forms of the stellar
IMF. While a Salpeter IMF is parametrised as usual $
\Phi(m)~\sim~m^{-2.35}$ over the range of stellar masses $m$ from $m_{\rm l} =
0.15~M_{\odot}$ to $m_{\rm u} \approx 85~M_{\odot}$, a Scalo IMF is used
in the form $\Phi(m)~\sim~m^{-x}$ with $x = -1.25$ for
$m \leq 1~M_{\odot}$, $x = -2.35$ for $1~M_{\odot}
< m \leq 2~M_{\odot}$, and $x = -3.00$ for $m > 2~M_{\odot}$. Normalization in both cases is $\int_{m_{\rm l}}^{m_{\rm u}} \Phi(m) \cdot m
\cdot dm = 1$. This implies that -- as compared to a Salpeter IMF -- the Scalo IMF has
more stars in the mass range $0.5~M_{\odot} \leq m \leq 6.5~M_{\odot}$
and less stars below $0.5~M_{\odot}$ and above $6.5~M_{\odot}$. 

\subsection{Input physics: Padova isochrones with TP-AGB}
The results from our evolutionary synthesis model are based on isochrones from
November, 18th, 1999 of the Padova group (for a description and discussion of
parameters for these isochrones with moderate core overshooting see Bertelli
\etal \cite{bert} and Girardi \etal \cite{girardi2}). Comparison of these
isochrones with CMDs of resolved star clusters in the Milky Way and LMC were
e.g. presented by Chiosi \etal (\cite{chiosi92}) and Vallenari \etal
(\cite{vallenari94}). We used the following data from their files: mass,
log$(T_{\rm E})$, $M_{\rm Bol}$, and $M_{\rm V}$ in their time
evolution. We do not use the colors and other luminosities also contained in
their files, because we obtain them from the spectra we calculate.

For the use with single short burst stellar populations, isochrones have a basic
advantage over the stellar evolutionary tracks they are based on. Because tracks
are only available for typically 50 (or less) stellar masses, drastic changes in
the luminosity evolution of an SSP occur when all stars of a certain mass
collectively climb on the red giant branch or die. Between reasonably defined
equivalent evolutionary stages of two stellar masses with evolutionary tracks
available, isochrones are interpolated in terms of
\begin{eqnarray*}
	\log{L}=\log{\left(L\left(m_{\rm i},\mu,\tau_\mu \right)\right)}\\
	\log{T_{\rm eff}}=\log{\left(T_{\rm eff}\left(m_{\rm i},\mu,\tau_\mu \right)\right)}
\end{eqnarray*}
where $m_{\rm i}$ is the initial mass for track i, $\mu$ the defined physical
phase of evolution and $\tau_\mu$ the relative duration of this phase.

It is then possible to interpolate both in luminosity and temperature between
two adjacent masses. For the interpolated masses the magnitudes and colors are
then obtained by translating the theoretical luminosities and effective
temperatures with tables of bolometric corrections and colors.

We supplemented these Padova isochrones by low mass stars in the mass range from
$0.15 - 0.45~M_{\odot}$ from the calculations of Chabrier \& Baraffe
(\cite{chab}) that include a new description of the interior of low mass objects
and use non-grey atmospheres (see Kurth \etal \cite{kurth} for
details). The Padova group recently added low mass-stars to their isochrones 
themselves (Girardi \etal
\cite{girardi2}) but did not compare their low mass stars to those of Chabrier \& Baraffe. The cumulative contribution of stars below $0.5~M_{\odot}$, however, to the integrated light is very small, anyway.

\begin{figure}[ht]
\includegraphics[width=\columnwidth]{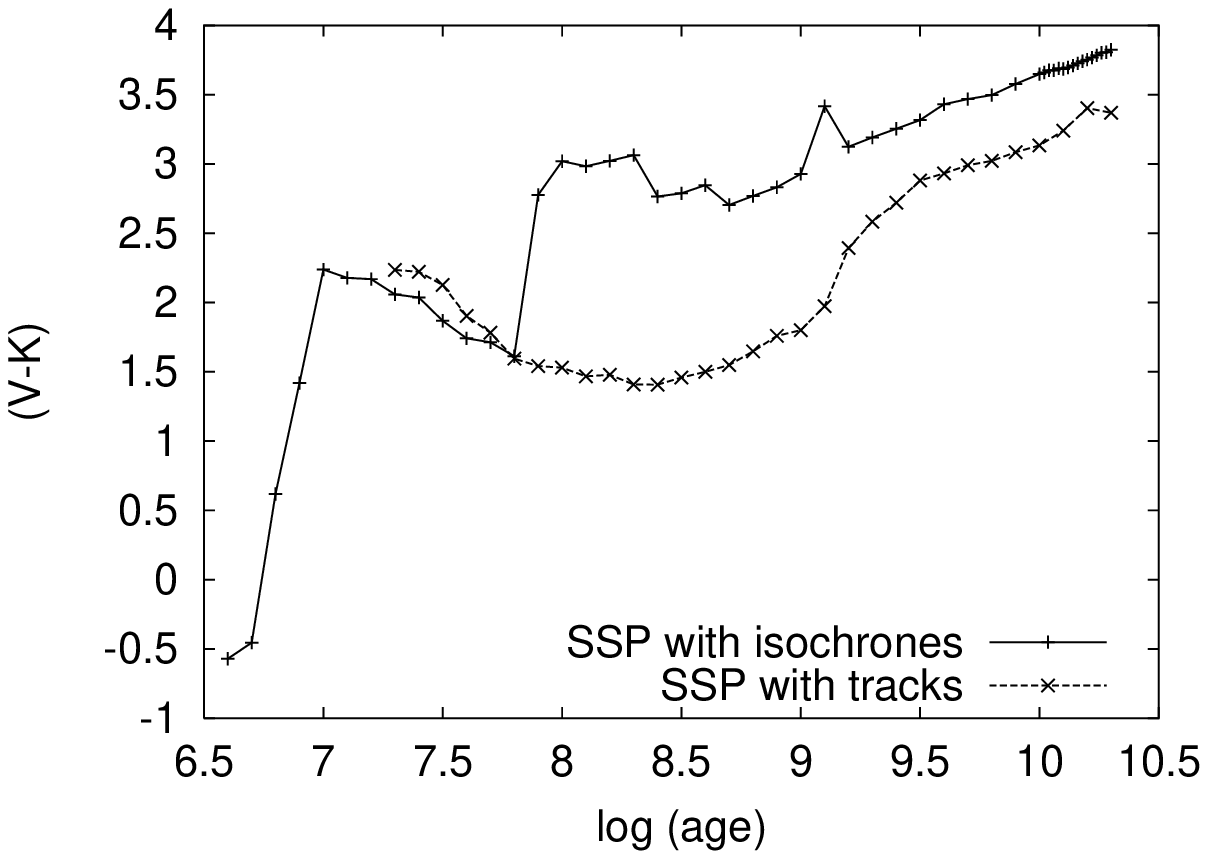}
\includegraphics[width=\columnwidth]{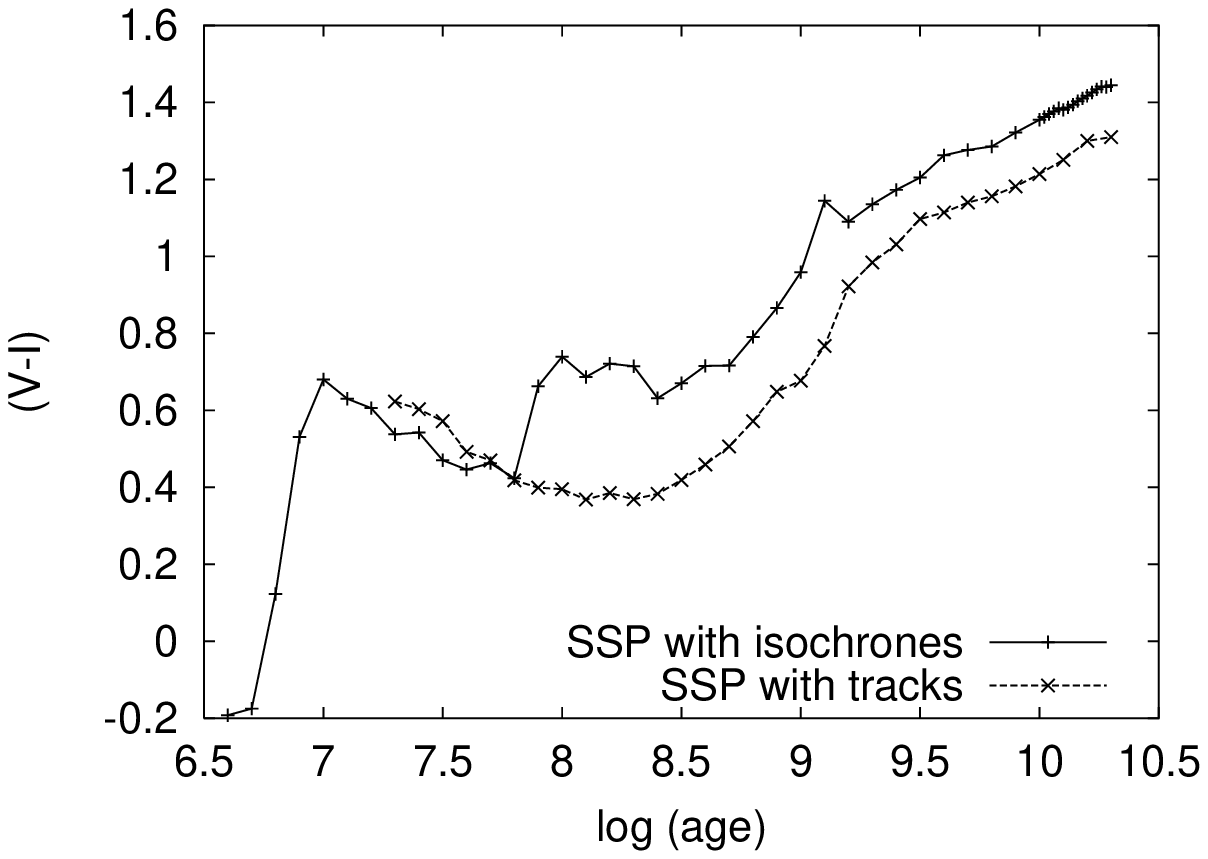}
\caption{Time evolution of $(V-K)$ and $(V-I)$ colors (as a function of logarithmic age [yr]) for SSPs 
based on isochrones, i.e. {\bf including the TP-AGB phase} (solid lines), and based on stellar evolutionary tracks that did not yet include the TP-AGB phase (dashed lines), both for solar metallicity.}
\label{iso-track}
\end{figure}

An important feature of the 1999 Padova isochrones is that they include a 
semi-analytical
description of the thermal pulsing asymptotic giant branch (TP-AGB) phase
(cf. Groenewegen \& de Jong \cite{tpagb}), which was not yet included in the
earlier Padova tracks (cf. Bressan \etal \cite{bree}, Fagotto \etal \cite{fagota}, \cite{fagotb}, \cite{fagotc}). 

TP-AGB stars are rare species in single burst populations, they are, however, very luminous and very cool (Lan\c{c}on \cite{lancon}). At ages between 0.1 and 1 Gyr, i.e. when the red supergiants have died already and the red giants are not important yet, TP-AGB stars account for 25 to 40 \% of the bolometric light of an SSP and for 40 to 60 \% of the light emitted in the $K-$band (cf. Charlot \cite{charlot}). 

Inclusion of the TP-AGB phase has strong effects on some of the colors. 
It makes the broad
band colors redder in the age range from $10^8$ to a few $10^9$
yr. This effect is strongest in $(V-K)$ and can amount to more than 1
mag in $(V-K)$ at solar metallicity as seen in Fig. \ref{iso-track},
where we plot the $(V-K)$ color evolution for an SSP based on
isochrones including the TP-AGB in comparison with an SSP from Kurth \etal
(\cite{kurth}) based on the Padova stellar evolutionary tracks that did not
include the TP-AGB phase. The effect is weaker at lower metallicities and for 
colors at shorter wavelengths. 

The effect of the TP-AGB phase has been discussed by the Padova group before
(Girardi \etal \cite{girardi1}) and the good agreement with $J-$ 
and $K-$ CMDs of resolved young open clusters is shown in Vallenari \etal
(\cite{vallenari99}). Rich \etal (\cite{rich01}) find excellent agreement of the
recent Padova isochrones with CMDs in $UBV$ and $I$ for the LMC cluster NGC 2121
which, at [Fe/H] $= -0.6$ and an age of 3.2 Gyr, is a very good test case
for the importance of the TP-AGB phase.

The effect of the TP-AGB phase is also significant in $(V-I)$ at
metallicities $\geq 0.5~Z_{\odot}$ and has important implications for the age
determination on the basis of colors observed for bright star clusters in
interacting and/or starburst galaxies. E.g. in our analysis of the sample of
young star clusters in the merger remnant NGC7252 which show a mean $
(V-I)=0.83 \pm 0.25$ we derived a mean age of $1.37 {+1.6 \atop -0.8}$ Gyr assuming a metallicity 
$\approx 0.5~ Z_{\odot}$ using SSP models based on Geneva stellar
evolutionary tracks (Fritze -- v. Alvensleben \& Burkert \cite{fb95}) very similar to what is found with the models of Kurth \etal
(\cite{kurth}) who use Padova tracks. Including the TP-AGB phase as in the
Padova isochrones used here, however, results in a mean age of $
8.7\cdot 10^8$ yr, i.e. only 64\% of the age obtained without the TP-AGB
phase. Most young star cluster systems observed with HST, in fact, have ages in the range
of $10^8$ to $10^9$ yr and metallicities $\geq 0.5~
Z_{\odot}$ where the effect of the TP-AGB phase on $(V-I)$ is
strong. Hence, all their age determinations on the basis of $(V-I)$ in
the literature are subject to this effect.

\subsection{Input physics: spectra}
We use the library of model atmosphere spectra from Lejeune \etal (\cite{lej2},
\cite{lej}) which is based on model atmospheres from Kurucz (see e.g. Kurucz
\cite{kurucz}), Fluks \etal (\cite{fluks}), and Bessell \etal (\cite{bessel89},
\cite{bessel91}). This library comprises the full range of metallicities covered
by the isochrones, is very complete in terms of stellar effective temperatures
from $T_{\rm eff}=2800~-~47500$ K, 
and gravities, and has a long wavelength coverage from 90 {\AA} to 160 $
\mu$m with good spectral resolution. Its spectra have partly been
corrected to provide good agreement with the observed colors of all types of
stars from $U$ through $K$ (cf. Lejeune \etal). Stars with an effective temperature
above 50000 K are described by pure black body spectra.

All the coolest AGB stars are long period variables which have different spectral type -- temperature -- color relations and, in particular, strongly increased molecular absorption features (e.g. $H_2O$ around 1.4 and 1.9 $\mu$m, $VO$ at 1.05 $\mu$m, . . . ) as compared to non-pulsating stars (Lan\c{c}on \cite{lancon}). Lan\c{c}on \& Wood (\cite{lw}) compiled an IR -- library of cool star spectra that was combined with the one of Lejeune \etal (\cite{lej}) that we use here and implemented into models by Fioc \& Rocca -- Volmerange (\cite{frv}). These models show that the inclusion of observed AGB spectra is very important for a description of specific spectral features in the near and mid-IR of intermediate age SSPs. Bressan \etal (\cite{bre98}) investigate in detail the effects of dust in the envelopes of AGB stars on the integrated IR properties of intermediate age SSPs. They consider both the absorption of light in the expanding dust shells of mass-losing stars at optical wavelengths and its thermal reemission in the IR. Differences with respect to a standard SSP without these dust absorption and emission features appear beyond a few $\mu$m where the contribution of the brightest AGB stars to the integrated light of the stellar population is large. Bressan \etal find a decrease of the flux between 1 and 3 $\mu$m because the brightest stars are heavily obscured by their circumstellar envelopes and thermal dust emission at longer wavelengths $\gta 4~\mu$m, reaching an order of magnitude at 10 $\mu$m over the pure photospheric models. 

The UV, optical and NIR regions we are studying in this paper are not affected by the expanding dust shells surrounding TP-AGB stars. 
We caution, however, that our models do not include the above-mentioned empirical AGB spectra yet, nor the absorption and reemission of light by their dusty envelopes, and, hence, should not be used for the interpretation of fluxes or spectral features of intermediate age star clusters beyond the $K-$band. In stellar populations with extended SFHs the AGB star features discussed above (molecular absorption bands, extinction and reemission by surrounding dust shells) are strongly overwhelmed by the emission of coexisting red supergiant and giant stars. 

\subsection{Isochrone synthesis model}
Our isochrone synthesis code for SSPs starts from a gas cloud of given mass and
metallicity and then forms a population of stars with a given SFR during the first
timestep of the calculation. It calculates a variety of observables at any
evolutionary timestep, typically $4 \cdot 10^6$ yr: 

\begin{itemize}
	\item masses in gas and stars
	\item ejection rates of gas and metals by dying stars
	\item spectra from 90 {\AA} ~to 160 $\mu$m
	\item luminosities, mass-to-light - ratios, and colors in various filters  
	\item evolutionary and cosmological corrections including the effect of 
	attenuation by intergalactic hydrogen. 
\end{itemize}

Ejection rates of gas and metals by dying stars are calculated from the individual stellar gas and heavy element yields for the respective isochrone metallicities. 

Every star of mass $m$ on an isochrone of age $t$ and metallicity $Z$ is attributed a spectrum ($S_{\ast}(\lambda,m)$) from
Lejeune \etal's spectral library, appropriate for its effective temperature
$T_{\rm eff}$ and surface gravity $g$. Because our model is a 1-zone model
without any spatial resolution, spectra of all stars along an isochrone of
given metallicity $Z$ can be summed up, weighted with the IMF $\Phi(m)$, 
to give the integrated isochrone spectrum ($S_{\rm iso}(\lambda,t,Z)$), i.e. the spectrum
of an SSP of metallicity $Z$ at age $t$
\begin{eqnarray}
	S_{\rm iso}(\lambda,t,Z)=\int_{m_{\rm l}}^{m_{\rm u}} S_{\ast}(\lambda,m,t,Z) \cdot \Phi(m) \cdot dm.
\label{eq:1}
\end{eqnarray}

Alternatively, the code can also model arbitrary stellar
populations with extended or complex star formation histories, such as galaxies. For the evolution of a galaxy with a star formation rate extended in time ($SFR(t)$), isochrones are followed as a
function of time and multiplied with the star formation rate at the formation time $t_{\rm i}$ of the respective isochrone, 
$SFR(t_{\rm i})$. 
So the integrated
spectrum of a galaxy ($S_{\rm gal}(\lambda,t,Z)$) at time t is calculated via Eq. (\ref{iso2}) 
\begin{eqnarray}
	S_{\rm gal} (\lambda,t,Z)=  \int^t_0 S_{\rm iso}(\lambda,(t - t_{\rm i}),Z) \cdot SFR (t_{\rm i}) \cdot dt_{\rm i}. 
\label{iso2}
\end{eqnarray}

To account for the fact that, in general, the stellar population of a galaxy is
comprised of subpopulations of different metallicities, an additional
integration over all the metallicities can be performed. Here, however, we only
present results for our SSP models.

\subsection{Filter systems }
Several filter systems are implemented in our evolutionary synthesis code, as
e.g. Johnson $U~B~V~R~I$ (cf. Lamla \cite{lamla}), $J~H~K~L$ (Bessel \& Brett
\cite{bessel}), $g~r$ (Thuan \& Gunn \cite{thuan}), $U^+~J^+~F^+~N^+$ (Kron \cite{kron}, Koo
\cite{koo}), HST $F300W$ . . . $F814W$, Washington $C~M~T1~T2$ (Harris \& Canterna \cite{wash}), and Str\"omgren $u~v~b~y$ (cf. Lamla \cite{lamla}). Results for all of
them are presented in our electronic tables.

Calibrations for the Washington filters were done with colors for stars in Gunn
\& Stryker's (\cite{stryker}) spectral library (Geisler \cite{geisler} {\sl
priv. comm.}). The Johnson, Thuan \& Gunn and Bessel \& Brett filter systems
were calibrated with information from Landolt--B\"ornstein VI/2b.  The
HST-Filters were calibrated with Kurucz's Vega model as described on the HST
home page. The calibration of Str\"omgren photometry was done according to Gray
(\cite{gray}).

With the time evolution of SSP spectra as presented here, any other filter
system within the long wavelength range of our spectra from 90 {\AA} through 160
$\mu$ m is readily applicable. Integrating over the spectra with
filter transmission and detector response functions allows to study both the
time evolution and the metallicity dependence of any arbitrary filter system.

\subsection{Cosmological models}
\label{sec:Cosmological-models}
As outlined in Sect. \ref{sec:Introduction}, our SSP results are not only
intended for use in the interpretation of star cluster data but also for
combination with any kind of dynamical galaxy formation and/or evolution
model. For convenient use in this second respect, we also calculate evolutionary
and cosmological corrections.

Spectra are redshifted to $\lambda' (z) = \lambda \cdot (1+z)$ and 
fluxes are decreased by $F'_{\lambda}(\lambda'(z),~z) =
F_{\lambda}(\lambda'(z),0)/(1 + z)$. 

For galaxies evolutionary and cosmological corrections are conventionally given
in terms of magnitude differences, which are normalized to the average
luminosity of locally observed galaxies of the respective type. SSPs, however,
do not have such an absolute magnitude scale. We therefore chose to give their
evolutionary and cosmological corrections in terms of luminosity ratios rather
than magnitude differences and call them $\epsilon_{\lambda}$ and $
\kappa_{\lambda}$, respectively, not to be confused with the $e_{\lambda}$
and $k_{\lambda}$ conventionally given for galaxies: 
\begin{eqnarray*}
\epsilon_\lambda = {L_\lambda (z, t(z)) \over L_\lambda(z,t_0)}\\
\kappa_\lambda = {L_\lambda (z,t_0) \over L_\lambda (0,t_0)},
\end{eqnarray*}
with $t(z)$ being the age of a galaxy observed at redshift $z$ and $t_0$ the age of a galaxy at $z=0$. 
Table \ref{bdm} gives
the ages t(z) of a stellar population formed at redshift $
z_{\rm f}=5.0$ and the {\bf B}olometric {\bf D}istance {\bf M}odulous ($BDM(z)$)
for two specific cosmological models $
(H_0,~\Omega_0,~\Lambda_0)~=~(75,~0.1,~0.9)$ and $(65,~0.1,~0)$ we chose
to present here.
With the large number of spectra given for every SSP($Z$) in short time intervals any arbitrary cosmological model can easily be applied. 

\begin{table}
\caption{Evolution time $t$ and $BDM$ as a function of redshift $z$ with the cosmology $H_0=75$, $\Omega_0=0.1$, $\Lambda_0=0.9$ on the left, and with 
$H_0=65$, $\Omega_0=0.1$, $\Lambda_0=0$ on the right hand side, respectively.}
\begin{tabular}[h]{cccp{1cm}ccc}
\hline
	$z$ &  $t$ [Gyr] & $BDM$   && $z$ & $t$ [Gyr] & $BDM$ \\
\hline 
	0.0 & $15.12$ & 36.512 && 0.0 & $11.92$ & 31.15\\
	0.5 & $9.932$ & 42.487 && 0.5 & $6.957$ & 42.267\\
	1.0 & $6.720$ & 44.375 && 1.0 & $4.535$ & 44.130\\
	1.5 & $4.632$ & 45.543 && 1.5 & $3.115$ & 45.305\\
	2.0 & $3.222$ & 46.377 && 2.0 & $2.189$ & 46.179\\
	2.5 & $2.236$ & 47.019 && 2.5 & $1.542$ & 46.877\\
	3.0 & $1.522$ & 47.536 && 3.0 & $1.065$ & 47.461\\
	3.5 & $0.988$ & 47.968 && 3.5 & $0.702$ & 47.961\\
	4.0 & $0.579$ & 48.337 && 4.0 & $0.417$ & 48.400\\
	4.5 & $0.258$ & 48.659 && 4.5 & $0.188$ & 48.790\\
	5.0 & $0.0$ & 48.945 && 5.0 & $0.0$ & 49.141\\
\hline
\end{tabular}

\label{bdm}
\end{table}

\section{Spectral Evolution of SSPs for 5 different metallicities}
\label{sec:Spectr-Evol-SSPs}
We present the time evolution of SSP spectra of 5 metallicities from $4 \cdot
10^6$ yrs (which is the youngest isochrone) up to 14 Gyr with a Salpeter and a
Scalo IMF, respectively. The initial stellar mass is $ 1.6 \cdot
10^9~M_{\odot}$ for both IMFs.

\subsection{Time evolution of spectra}

\begin{figure}[ht]
\includegraphics[width=\columnwidth]{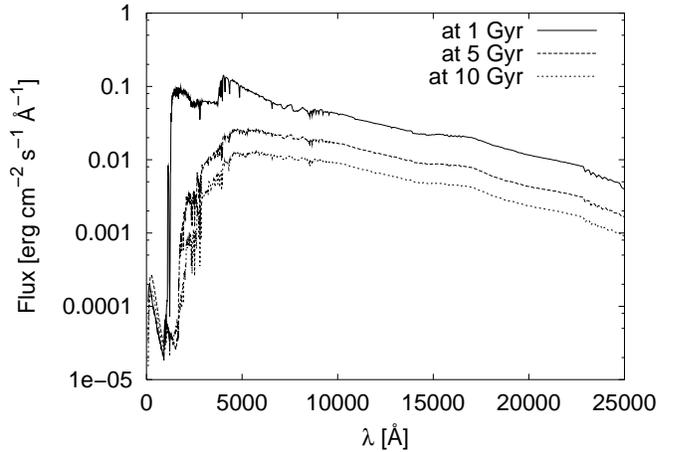}
\caption{Spectrum in terms of flux $F_{\lambda}$ as a function of wavelength $\lambda$ at 3 different
times for an SSP of solar metallicity and Salpeter IMF.}
\label{spec-age_sp}
\end{figure}

In Fig. \ref{spec-age_sp} we present the spectral evolution of an SSP with solar
metallicity and Salpeter IMF at 3 different ages of 1, 5, and 10 Gyr,
respectively. The flux decreases with time as stars of lower and lower masses
die with the strongest effect seen between 1500 and 5000 {\AA}. Note the small
rise of the flux below 1000 {\AA} due to very hot stars on the white dwarf
cooling sequence.

The 1 Gyr spectrum shows clearly the Balmer, Paschen, and Bracket series of
hydrogen absorption lines. Note the almost unchanging flux at $\lambda \lta
1000$ {\AA} and the strong time evolution in the 4000 {\AA} break.

The spectral evolution of an SSP with a Scalo IMF looks very similar to the one with a Salpeter IMF plotted here, because, at 1 Gyr, most of the high mass stars that were less numerous in an SSP with Scalo IMF, have already
died. A small difference in absolute flux would be seen due to the higher 
number of low mass stars in the Scalo IMF, which dominate the flux at higher ages.

\subsection{Metallicity effects on the spectral evolution}

\begin{figure}[ht]
\includegraphics[width=\columnwidth]{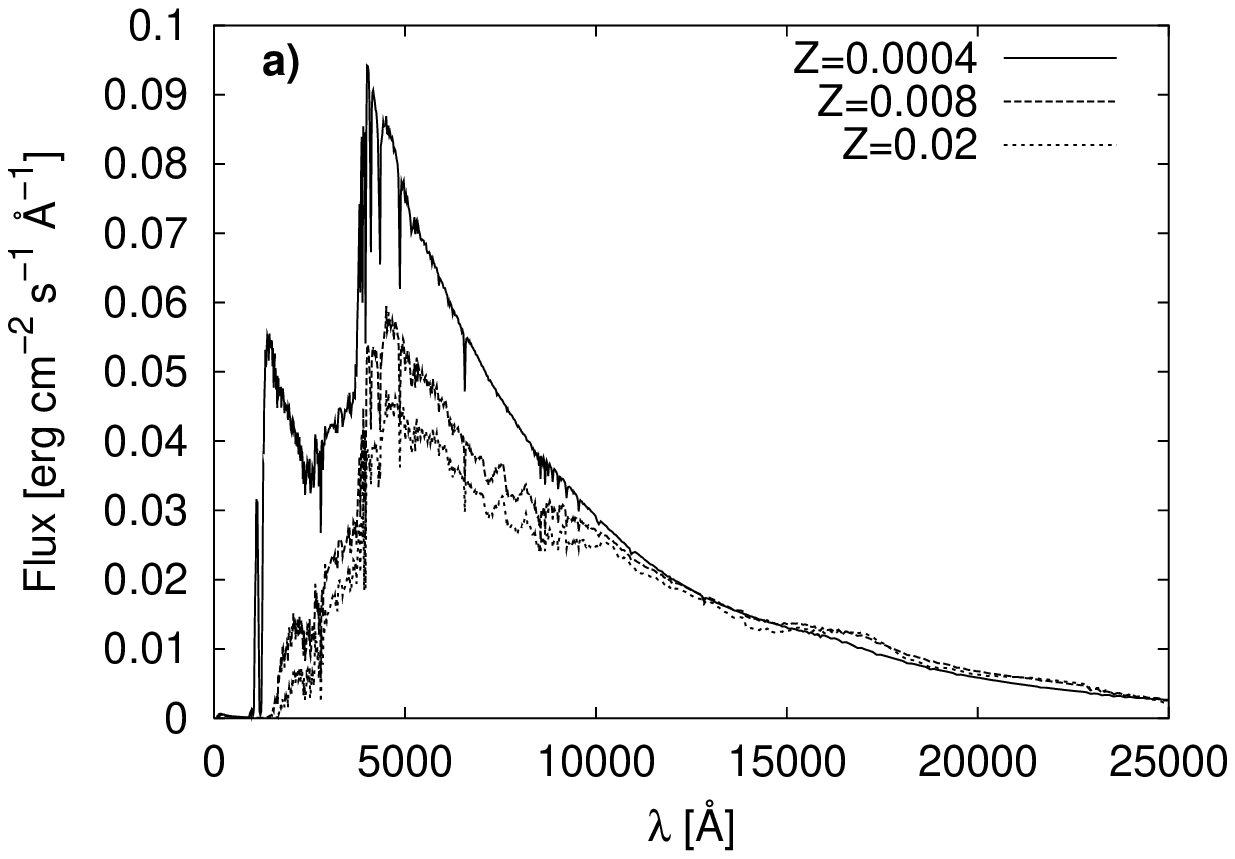}
\includegraphics[width=\columnwidth]{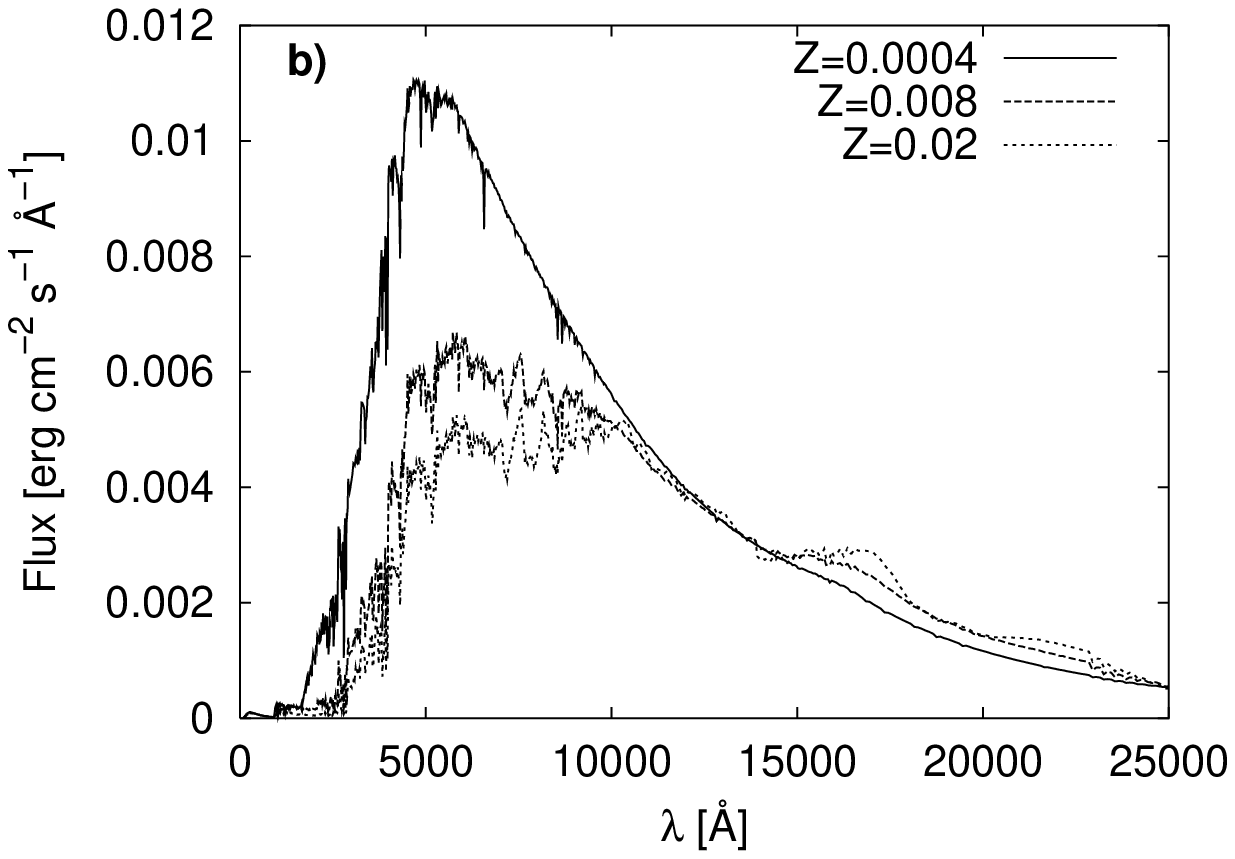}
\caption{Flux $F_{\lambda}$ as a function of wavelength for SSPs with Salpeter 
IMF and 3 different metallicities at ages of {\bf a)} 1 Gyr and {\bf b)} 10 Gyr. ($Z=0.0004=0.02~Z_{\odot},~Z=0.008=0.4~Z_{\odot},~Z=0.02=Z_{\odot}$).}
\label{spec-z_sp}
\end{figure}

In Figs. \ref{spec-z_sp} a) and b), we show spectra for SSPs of 3 different
metallicities at ages of 1 and 10 Gyr, respectively, all with a Salpeter IMF.
Enormous differences are seen both at young and old ages between SSPs of
different metallicities. As compared to a solar metallicity SSP, the
spectrum of a low metallicity SSP shows significant differences both in
continuum slope and absorption line strengths. Over the $UV$ and optical
wavelength range the continuum spectrum of a low metallicity SSP has stronger
flux and the maximum at shorter wavelengths as compared to a solar metallicity
SSP. This is a consequence of stars -- on average -- being brighter and hotter
at lower metallicity.  Note the very
strong $UV$ flux at $1000$ {\AA} $\lta \lambda \lta 2000$ {\AA} of the 1 Gyr very
low metallicity ($Z = 0.02~Z_{\odot}$) SSP. Not only the metal absorption lines
are weaker -- as expected -- in low metallicity SSPs of all ages, but also the
Balmer absorption lines, most prominent in the 1 Gyr old SSPs, are visibly
stronger at low metallicities (cf. Kurth \etal \cite{kurth} for a quantitative
analysis of absorption features).

At young ages $\sim 1$ Gyr, the near infrared (NIR) continuum is largely
metallicity independent while at older ages, e.g. at 10 Gyr as shown in
Fig. \ref{spec-z_sp}b, it is visibly affected by metallicity effects. E.g. in
the 1.7 $\mu$m bump in the $H-$band, and at 2.2 $\mu$m 
in the $K-$band a 10 Gyr old $Z=0.02~Z_{\odot}$ SSP only shows $\sim
\frac{2}{3}$ of the flux of a solar metallicity SSP.

\subsection{Comparison with earlier models}
\begin{figure}[ht]
\includegraphics[width=\columnwidth]{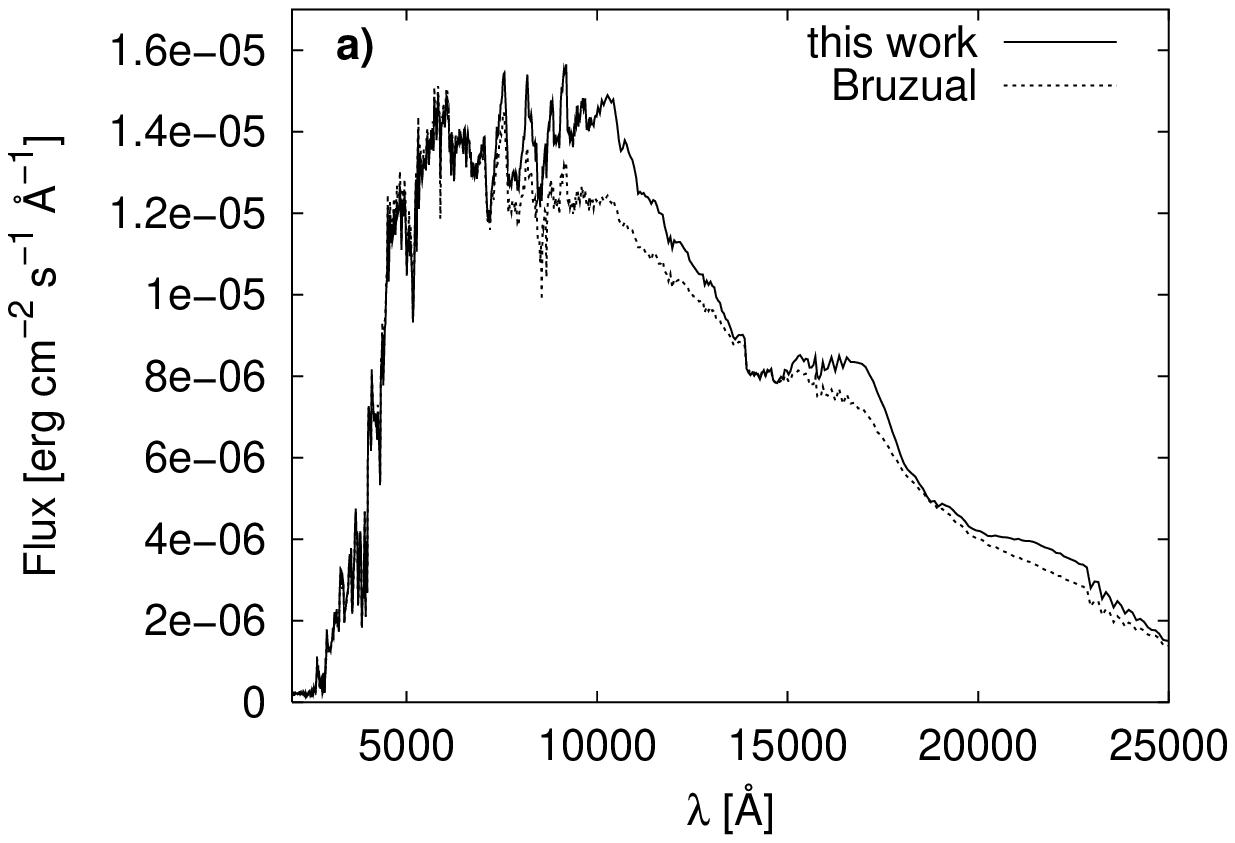}
\includegraphics[width=\columnwidth]{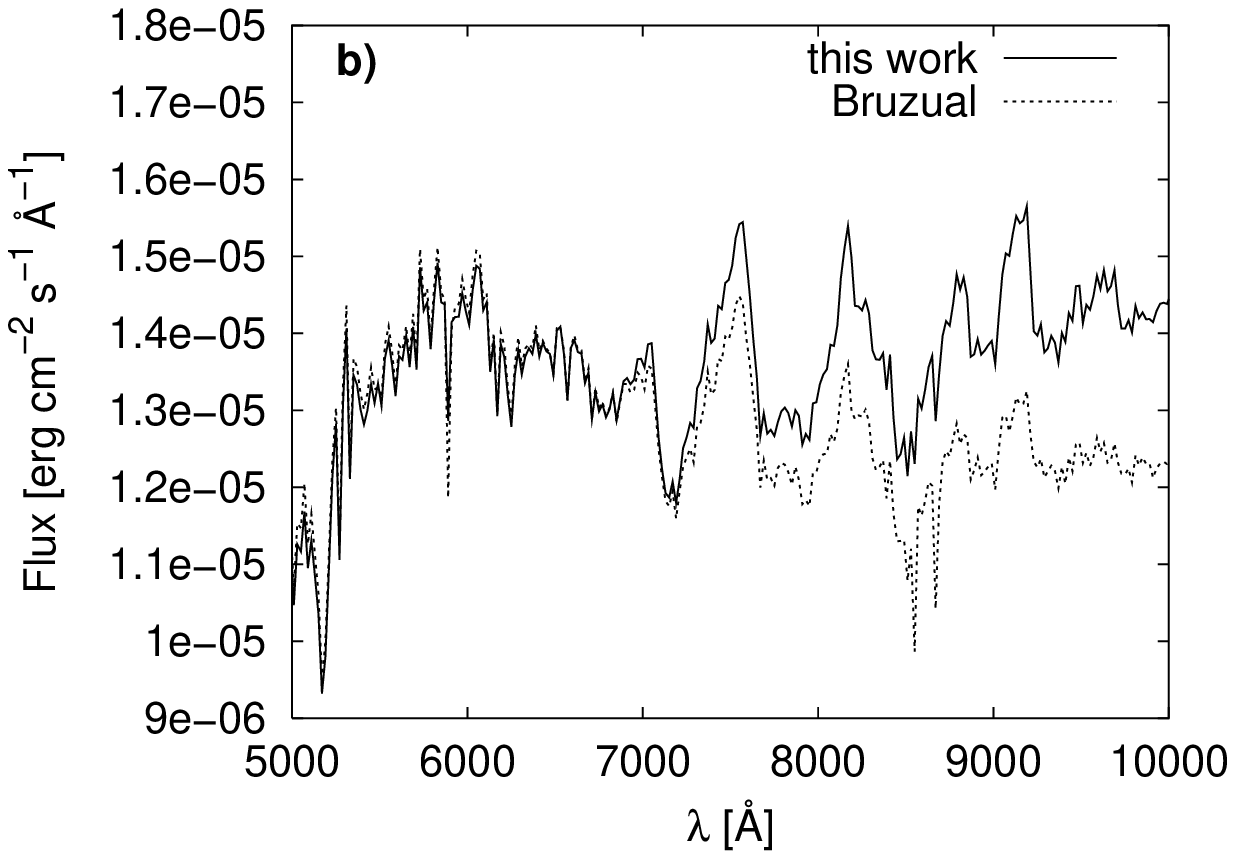}
\includegraphics[width=\columnwidth]{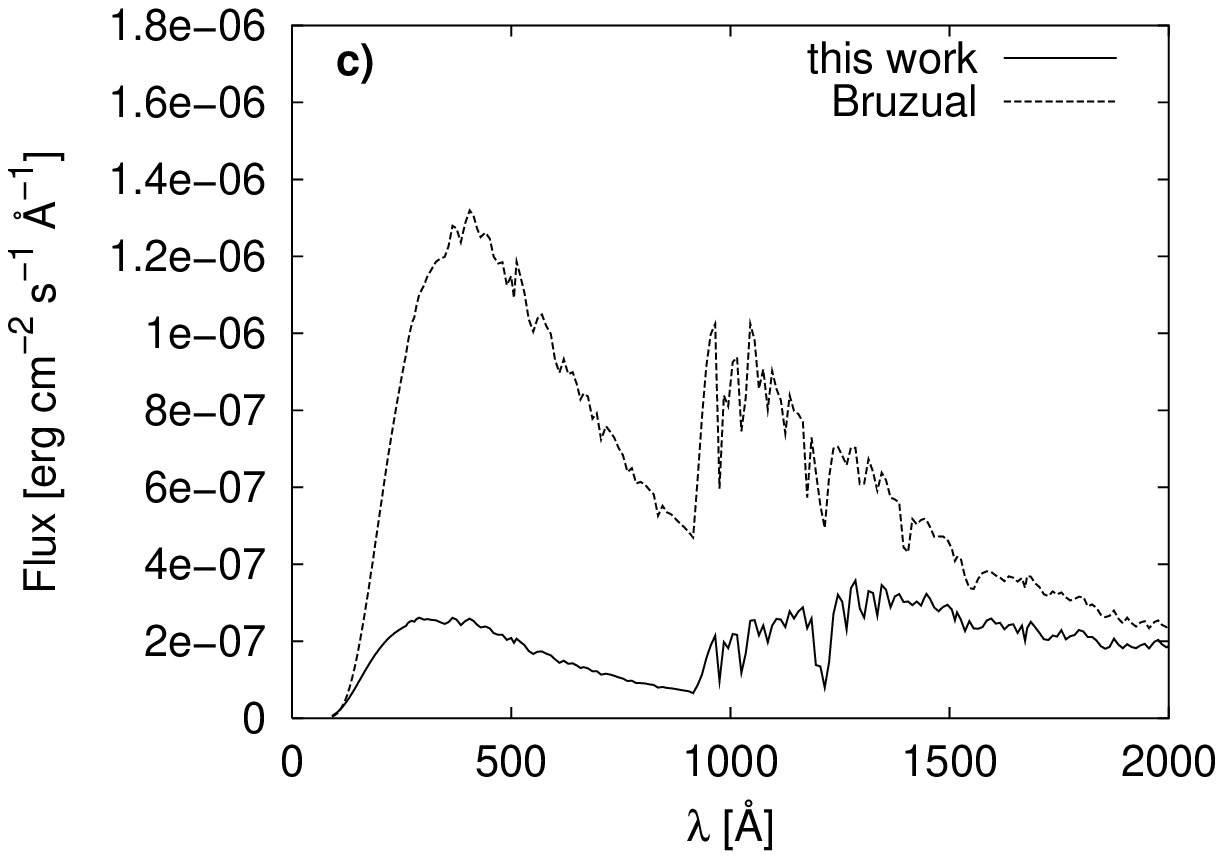}
\caption{Spectra in terms of flux as a function of wavelength. Comparison between our
model and the model from Bruzual \& Charlot at 12 Gyr with Salpeter IMF and solar metallicity
in 3 different wavelength ranges.}
\label{spec-ssp_b}
\end{figure}

We compare our results with those from Bruzual \& Charlot's (\cite{bruz2})
models (hereafter {\bf BC}) as obtained from the CD-ROM accompanying Leitherer
\etal (\cite{bruz}).  In Fig. \ref{spec-ssp_b} we show a comparison of spectra
from 1000 {\AA} to 25000 {\AA} a), a more detailed view from 5000 {\AA} to 10000
{\AA} b) and a view from 0 to 2000 {\AA} c) for SSPs with solar metallicity and
Salpeter IMF.  Both models show good agreement from the $UV$ to $\sim 7000$
{\AA}. Between 7000 {\AA} and 12000 {\AA}, however, BC's spectrum shows
significantly less flux than ours. The same happens again in the $H-$ and $K-$band
regions. This probably is due to the TP-AGB phase which involves stars with an
effective temperature around 3000 K.  The comparison in the wavelength range
from 90 {\AA} to 2000 {\AA} shows a strong difference between the models. In
this wavelength region very hot stars dominate the flux. These are
stars on the white dwarf cooling sequence. We do not know in detail
how theses stars are handled in BC's models but we suspect that the flux
difference is due to a different treatment of white dwarfs.

\section{Luminosity Evolution, Colors, and Mass-to-Light Ratios}
\label{sec:Lumin-Evol-Colors}
\subsection{Luminosity evolution}
\begin{figure}[ht]
\includegraphics[width=\columnwidth]{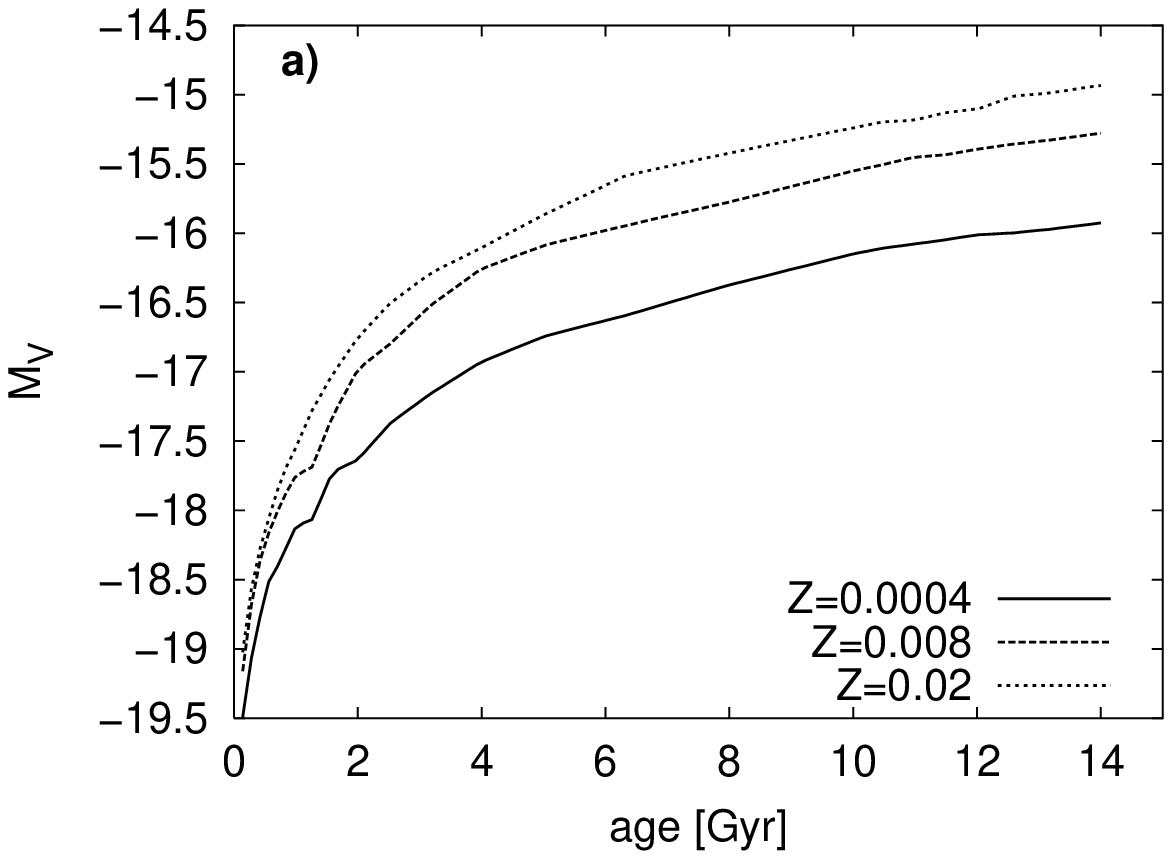}
\includegraphics[width=\columnwidth]{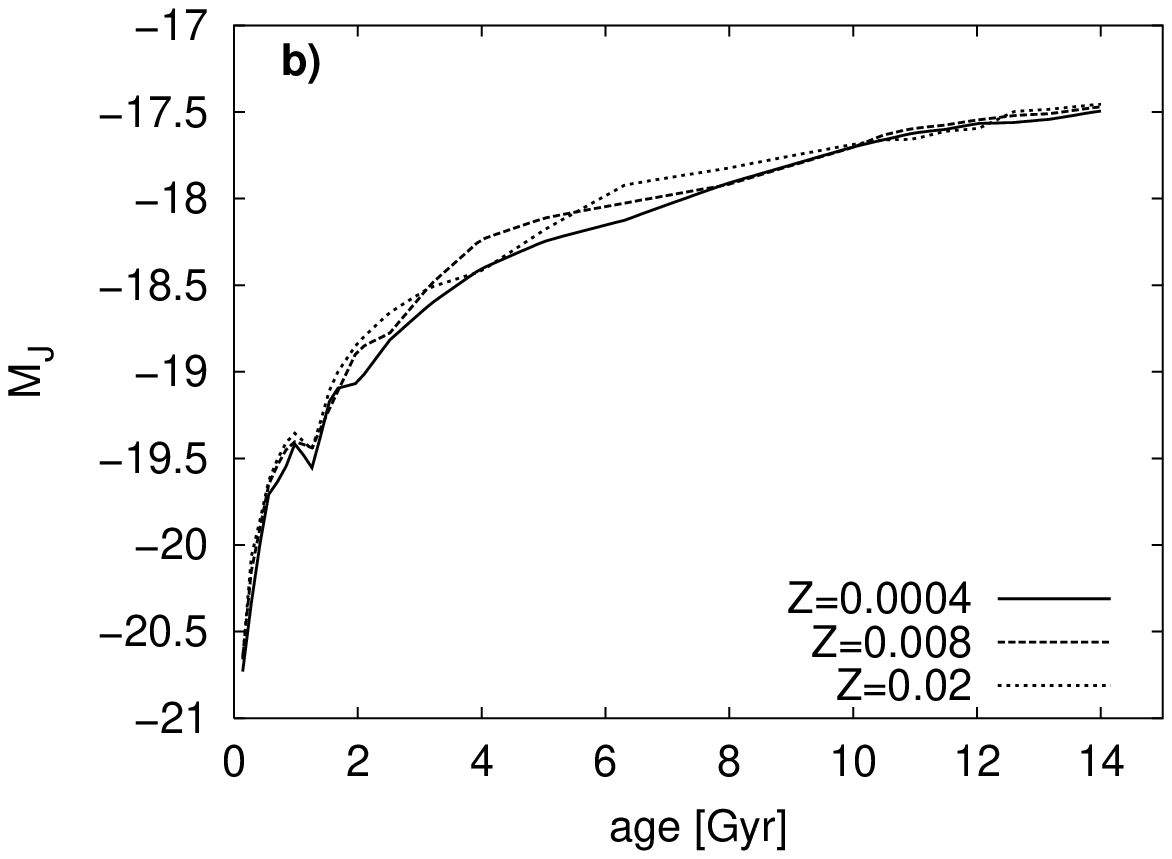}
\includegraphics[width=\columnwidth]{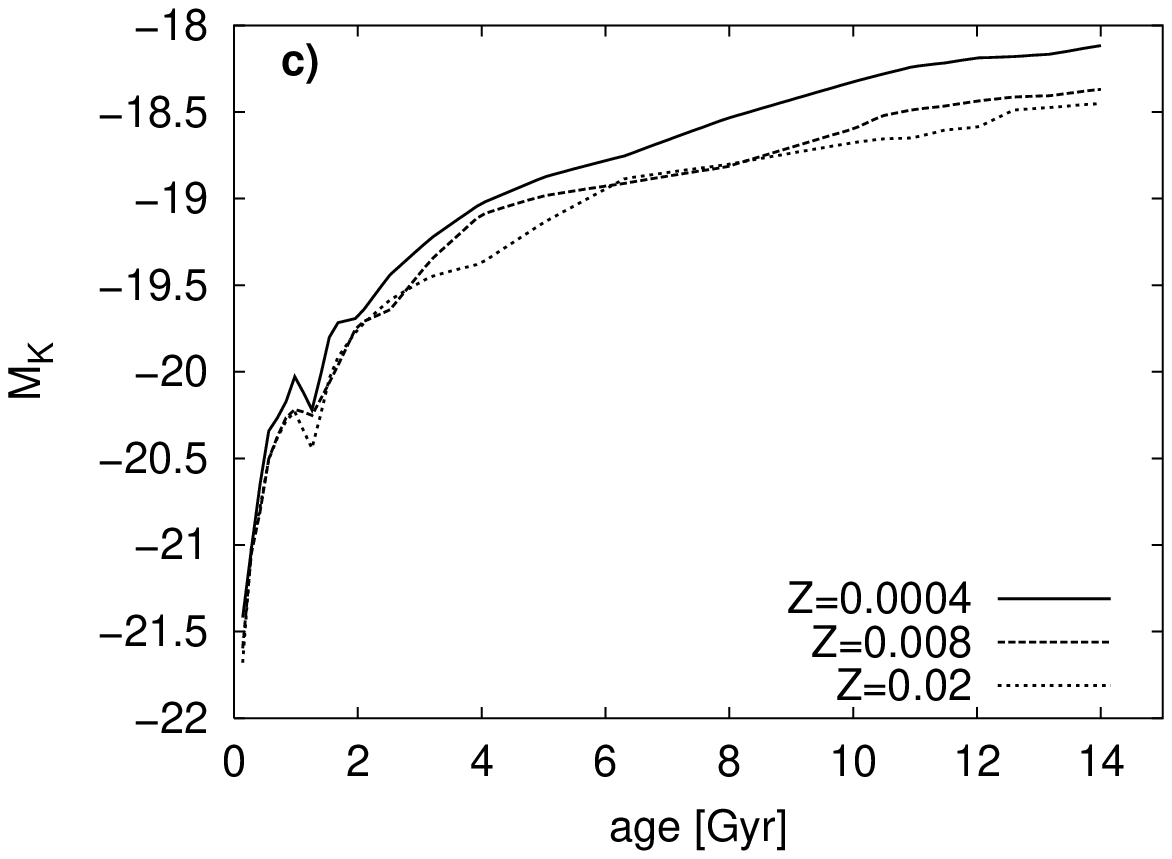}
\caption{Time evolution of absolute magnitudes in Johnson $V-$ ({\bf a}), $J-$ ({\bf b}), and $K-$bands ({\bf c}) for SSPs with Salpeter IMF and 3 different metallicities. 
($Z=0.0004=0.02~Z_{\odot},~Z=0.008=0.4~Z_{\odot},~Z=0.02=Z_{\odot}$).}
\label{m_sp}
\end{figure}

In Fig. \ref{m_sp} we present the time evolution of $V-,~ J-$, and $K-$band absolute
magnitudes of SSPs with 3 different metallicities. Again, the metallicity
effects are discussed on the example of a Salpeter IMF.

All over the time evolution, SSPs are brighter at lower metallicities in all bands $U$, \dots, $I$ (Fig. \ref{m_sp}a). The difference in
$M_V$ between a solar metallicity SSP and one with $0.02~Z_{\odot}$ increases from $\lta 0.5$ mag at few $10^8$ yr to $\sim 1$ mag at 5
-- 14 Gyr.

As seen in Fig. \ref{m_sp}b, the time evolution of $M_J$ does not show
any significant metallicity dependence, as could already be expected from the
spectra in Figs. \ref{spec-z_sp}a, b. The spectra also indicate what is clearly
found for the $H-$ and $K-$band evolution: the trend of luminosity with metallicity
is reversed in the near infrared spectral region as compared to optical
wavelengths with the higher metallicity SSPs being brighter in $H$ and $K$
(cf. Fig. \ref{m_sp}c), although the luminosity differences are not as strong as they are at shorter wavelengths.

\subsection{Color evolution}
\begin{figure}[ht]
\includegraphics[width=\columnwidth]{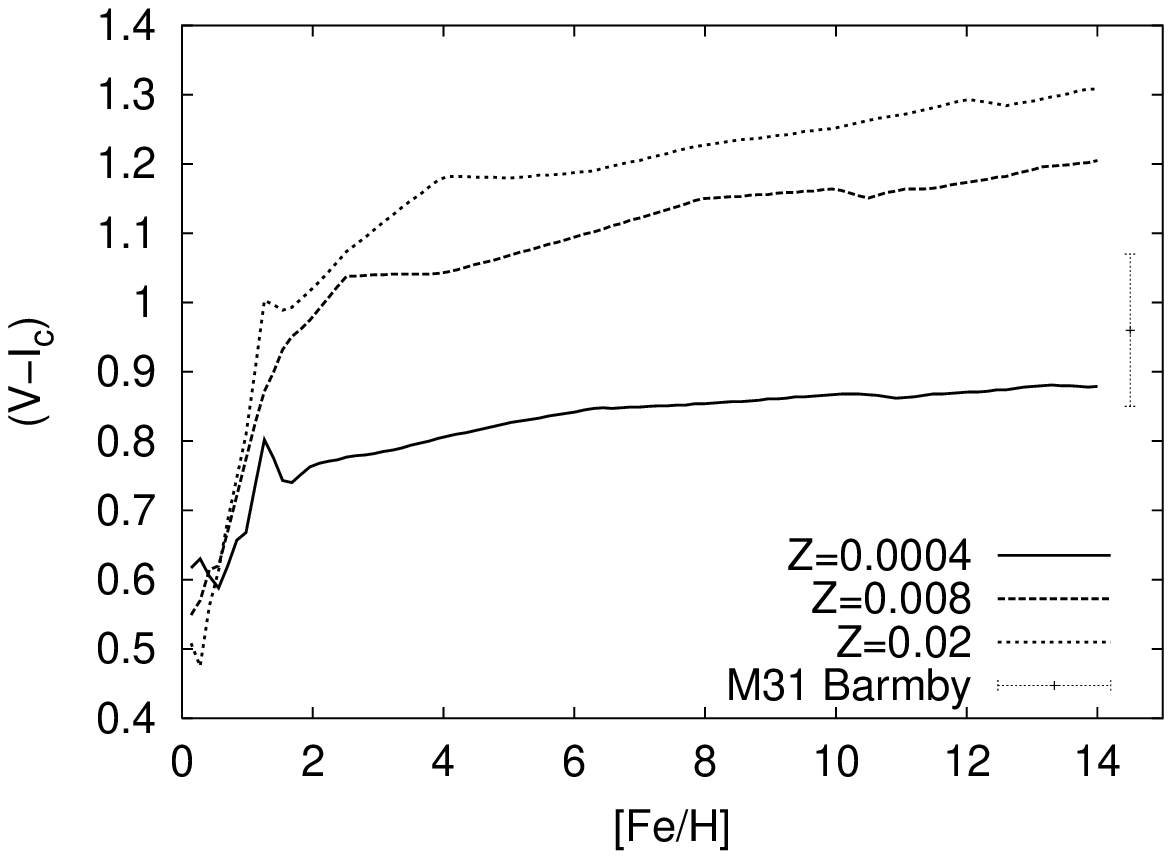}
\includegraphics[width=\columnwidth]{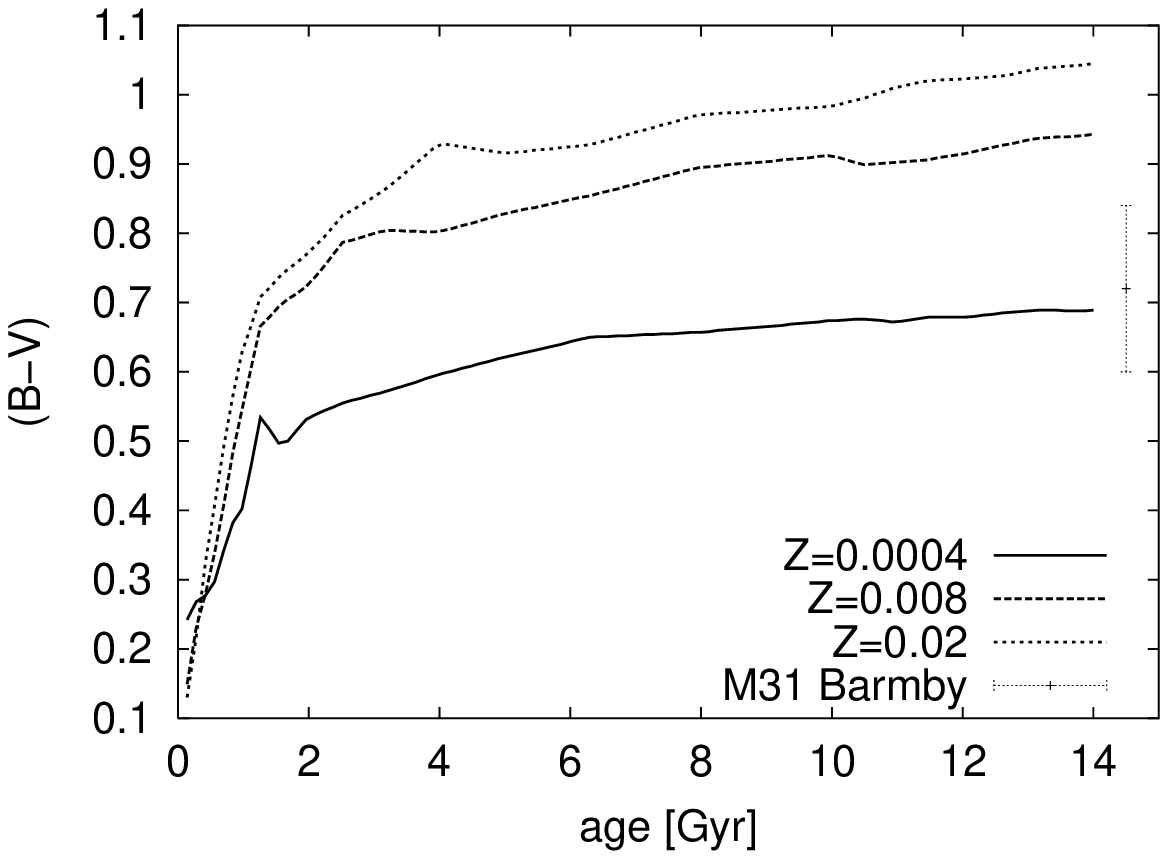}
\caption{Color evolution $(V-I_{\rm c})$ ($I_{\rm c}$ is Cousins I) for SSPs with
Salpeter IMF and three different metallicities. On the right we show the
mean color of M31 GCs (with $1\sigma$ range) as given by Barmby \etal. 
($Z=0.0004=0.02~Z_{\odot},~Z=0.008=0.4~Z_{\odot},~Z=0.02=Z_{\odot}$).}
\label{v-i_sp}
\end{figure}

The $(V-I_{\rm c})$ and $(B-V)$ color evolution for SSPs of
various metallicities with a Salpeter IMF is plotted in Fig. \ref{v-i_sp}. Note
that $I_{\rm c}$ now is the Cousins $I$ for the purpose of comparison
with data for M31 clusters. On the right hand side of Fig. \ref{v-i_sp} we give
the median colors with their 1 $\sigma$ dispersions of M31 clusters which have a mean metallicity [Fe/H] $= -1.2$ as determined by Barmby \etal (\cite{barmby}). We note that observed integrated GC colors are in good
agreement with our models for the corresponding metallicity at ages 12 -- 14
Gyr.

During the first $\sim 1$ Gyr, $(V - I_{\rm c})$ colors of all three SSPs
show a complex behavior with the solar metallicity SSP starting out $\sim
0.1$ mag bluer and reddening much more than the one with $0.02~Z_{\odot}$. For
all ages $\gta 1 - 2$ Gyr, however, SSPs of different metallicities show largely
different $(V-I_{\rm c})$ colors. E.g., at an age of 12 Gyr, the SSPs with $Z_{\odot},~0.4~Z_{\odot}$ and $0.02~Z_{\odot}$ have $(V-I_{\rm c})=1.3,~1.17$, and $0.87$, 
respectively. Hence, if e.g. star cluster ages
are known to be $\gta 2$ Gyr, their observed $(V-I_{\rm c})$ colors allow
for largely unambiguous metallicity determinations.

Note that differences between the two IMFs are very small, as already mentioned
above, and therefore not shown here. In particular, they almost do not affect age
and/or metallicity determinations at ages ${\rm \gta 4 \cdot 10^8}$ yr.

\subsection{Mass-to-light ratios}
\begin{figure}[ht]
\includegraphics[width=\columnwidth]{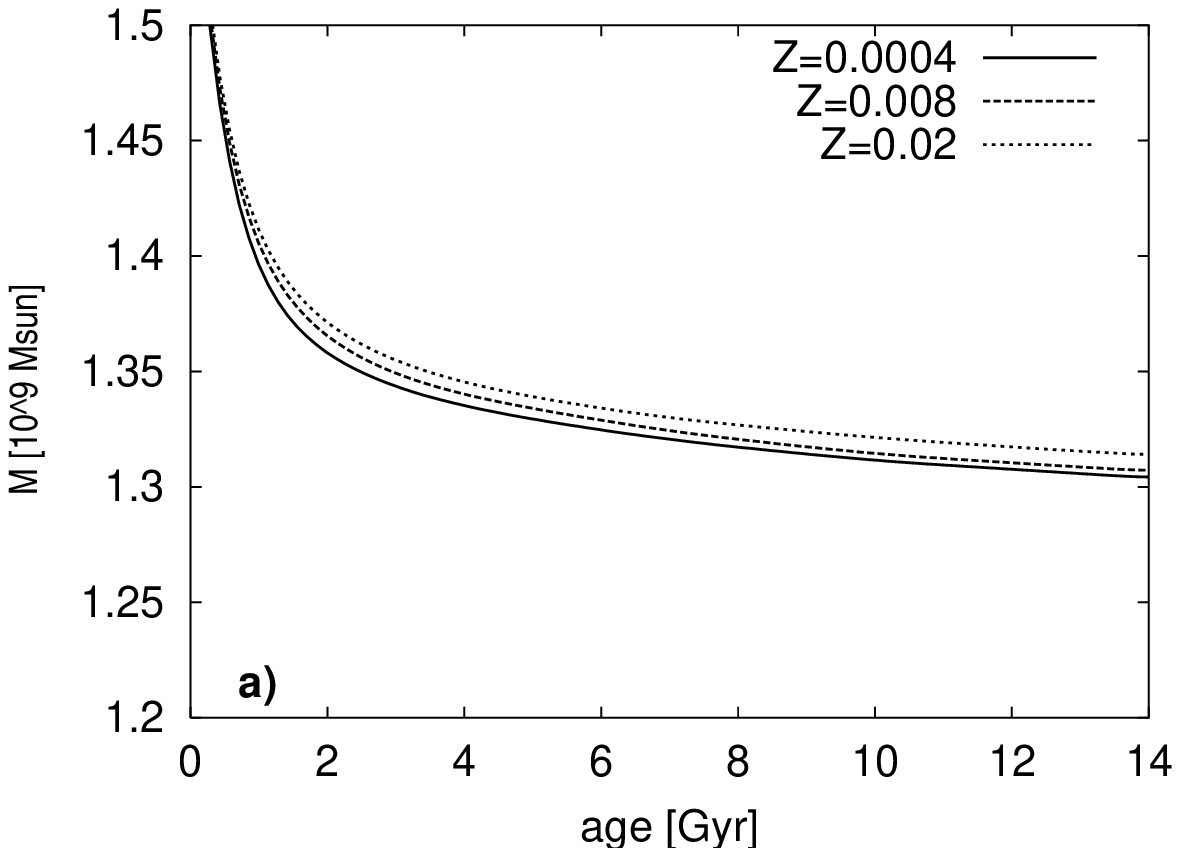}
\includegraphics[width=\columnwidth]{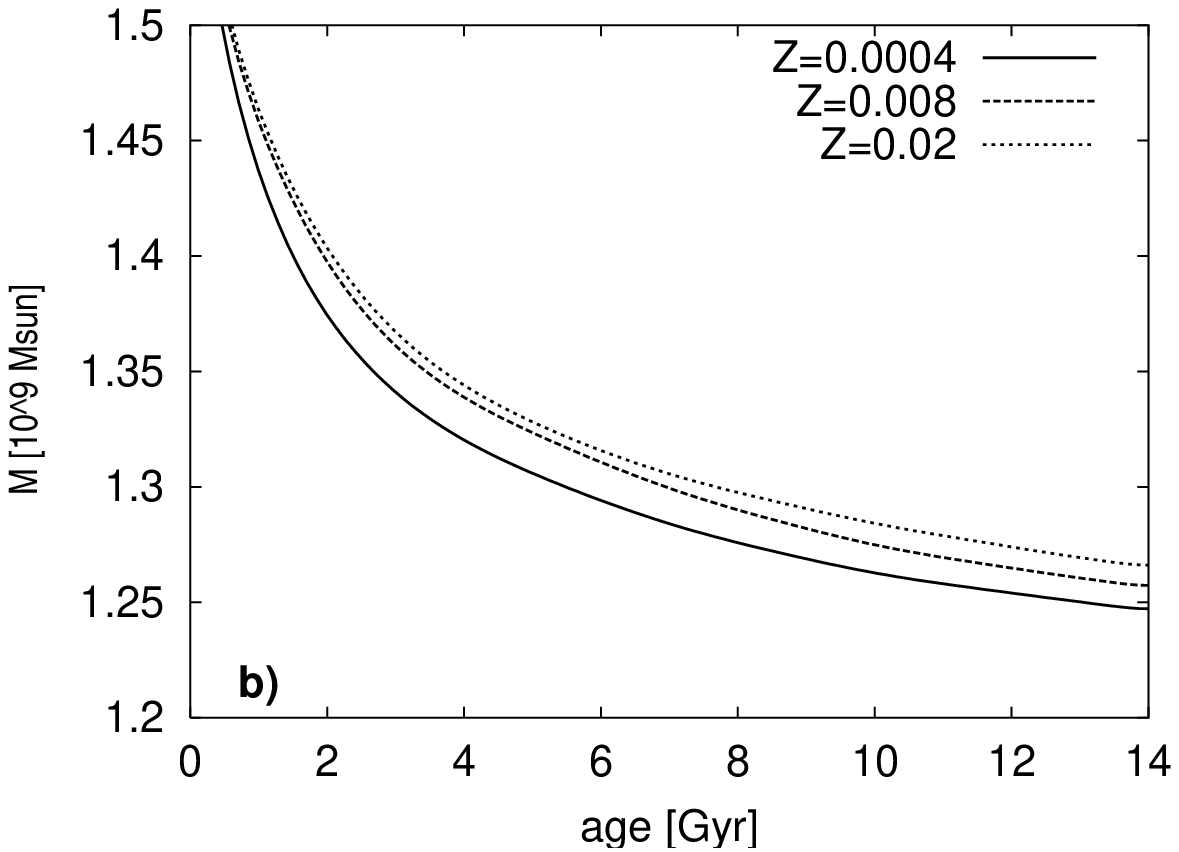}
\caption{Time evolution of the total stellar mass $M$ in units of $10^9~M_{\odot}$ for SSPs with {\bf a)} Salpeter and {\bf b)} Scalo IMF and three
different metallicities. 
($Z=0.0004=0.02~Z_{\odot},~Z=0.008=0.4~Z_{\odot},~Z=0.02=Z_{\odot}$).}
\label{sm_sp}
\end{figure}

In Fig. \ref{sm_sp} we show the time evolution of the total mass in stars, $M$, in SSPs with Salpeter and Scalo IMFs and three different metallicities. In all
cases, SSPs started out with an initial stellar mass $M = 1.6 \cdot
10^9~M_{\odot}$ as mentioned above.

The total stellar mass $M$ decreases as stars lose mass by stellar winds and die after
ejecting mass either in a PN or in a SN event. With a Salpeter and a Scalo IMF,
SSPs lose $\sim 19~\%$ and $\sim 23~\%$, respectively, of their initial mass
until an age of 12 -- 14 Gyr. The bulk of the mass loss occurs during the first
$\sim 2$ Gyr in case of a Salpeter IMF due to its larger number of high mass
stars, while mass loss occurs more steadily with a Scalo IMF.

While for an SSP with Salpeter IMF the metallicity dependence of the mass loss
is very small, it is slightly stronger in the case of a Scalo IMF (23\% for
$Z_{\odot}$ and 20\% for $0.02~Z_{\odot}$).

Although the difference in mass loss is small between SSPs with Salpeter and
Scalo IMFs, the differences in mass-to-light ratios $M/L_{\lambda}$,
with $\lambda$ denoting any filter band, are much larger because of the
luminosity differences.

\begin{figure}[ht]
\includegraphics[width=\columnwidth]{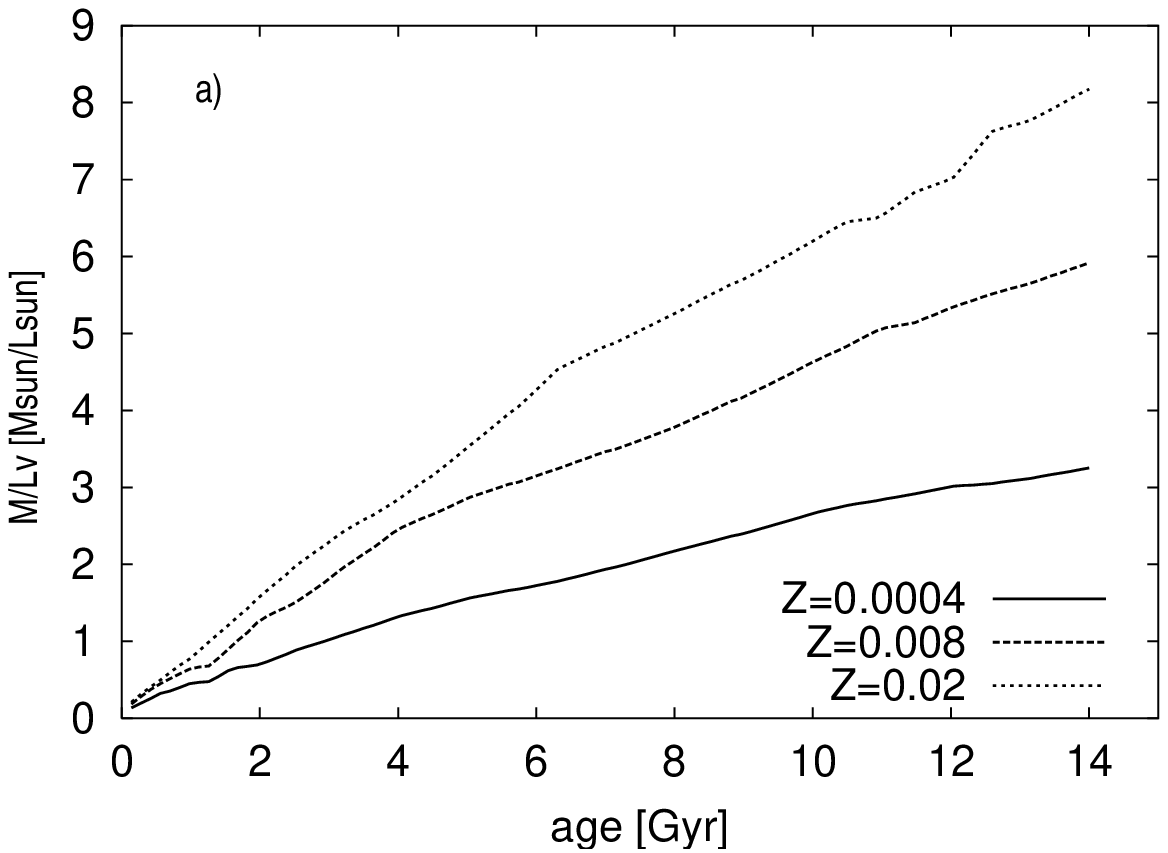}
\includegraphics[width=\columnwidth]{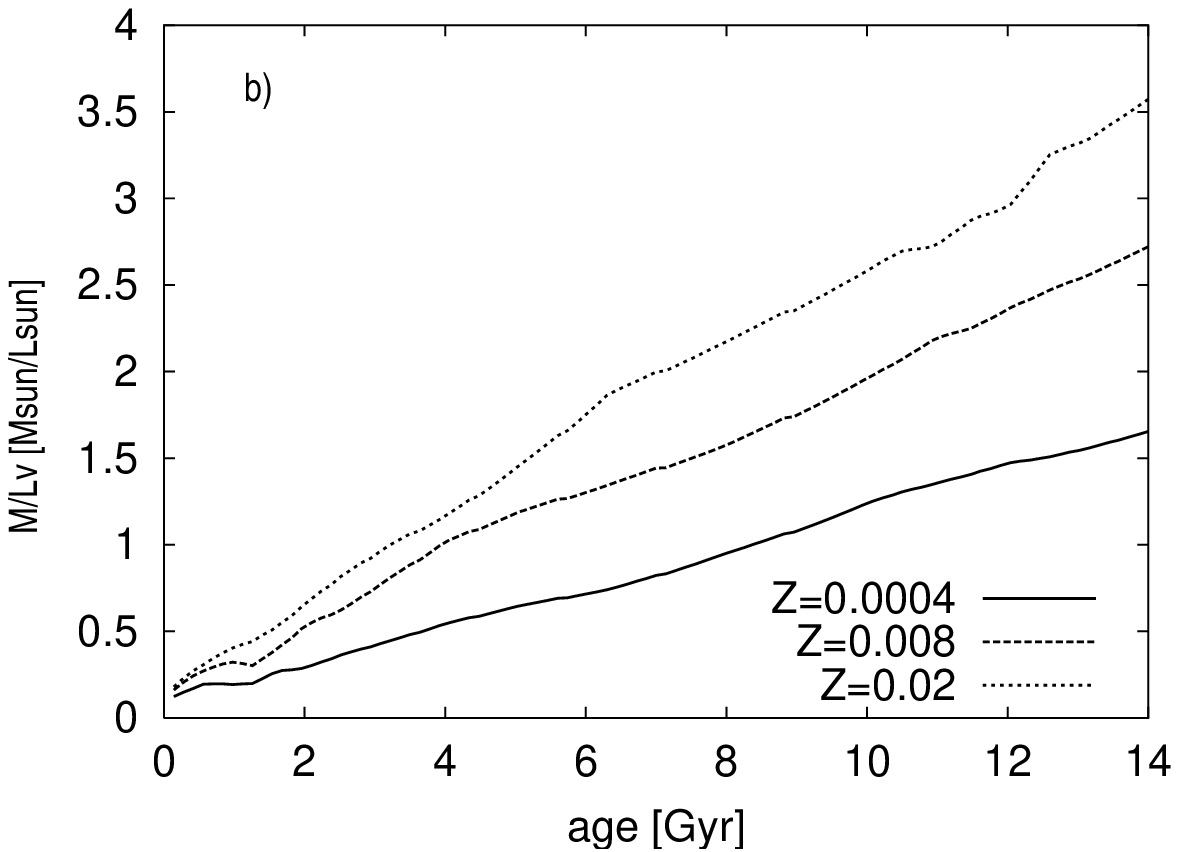}
\caption{Time evolution of $M/L_{\rm V}$ in solar units for SSPs with Salpeter (top panel) and
Scalo (bottom panel) IMFs and three different metallicities. 
($Z=0.0004=0.02~Z_{\odot},~Z=0.008=0.4~Z_{\odot},~Z=0.02=Z_{\odot}$).}
\label{m_over_l_sp}
\end{figure}

Fig. \ref{m_over_l_sp} shows the time evolution of $V-$band mass-to-light ratios
$M/L_{\rm V}$ of SSPs with both IMFs and three metallicities. The mass is the
time evolving stellar mass of the SSP with mass loss included as depicted in
Fig.\ref{sm_sp}.

Note the difference of more than a factor of 2 in $M/L_{\rm V}$ at all times
and for all metallicities between a Salpeter and a Scalo IMF and also the strong metallicity dependence in both cases.

At 10 Gyr, a Salpeter SSP has $M/L_{\rm V} \sim 12,~\sim 9,~\sim 5$ in solar
units for $Z = Z_{\odot},~ 0.4~Z_{\odot}$, and $0.02~Z_{\odot}$, respectively. For an
extensive discussion of $M/L-$values (and their uncertainties) both theoretically
derived by various groups using different sets of input physics and in
comparison with observationally derived $M/L-$values obtained for GCs using
various methods see Fritze - v. Alvensleben (\cite{fritze}).

\section{Theoretical calibrations of colors in terms of metallicity and
comparison with GC data} 
\label{sec:Theor-calibr-colors}
When spectroscopic abundance determinations are inaccessible or
exceedingly time-consuming, like e.g. for GCs in external galaxies, broad band
colors in combination with empirical calibrations with well known abundances are used (cf. Couture \etal \cite{cort}).

\begin{figure}[ht]
\includegraphics[width=\columnwidth]{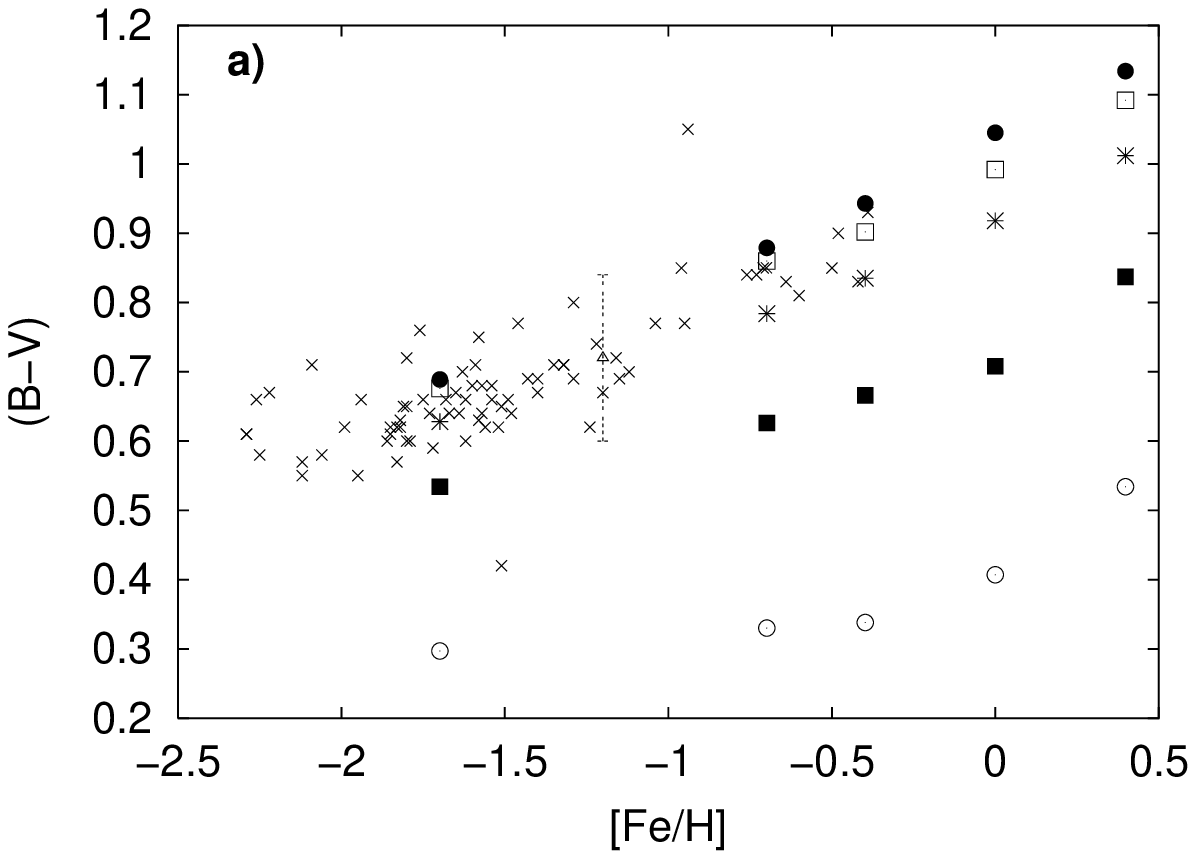}
\includegraphics[width=\columnwidth]{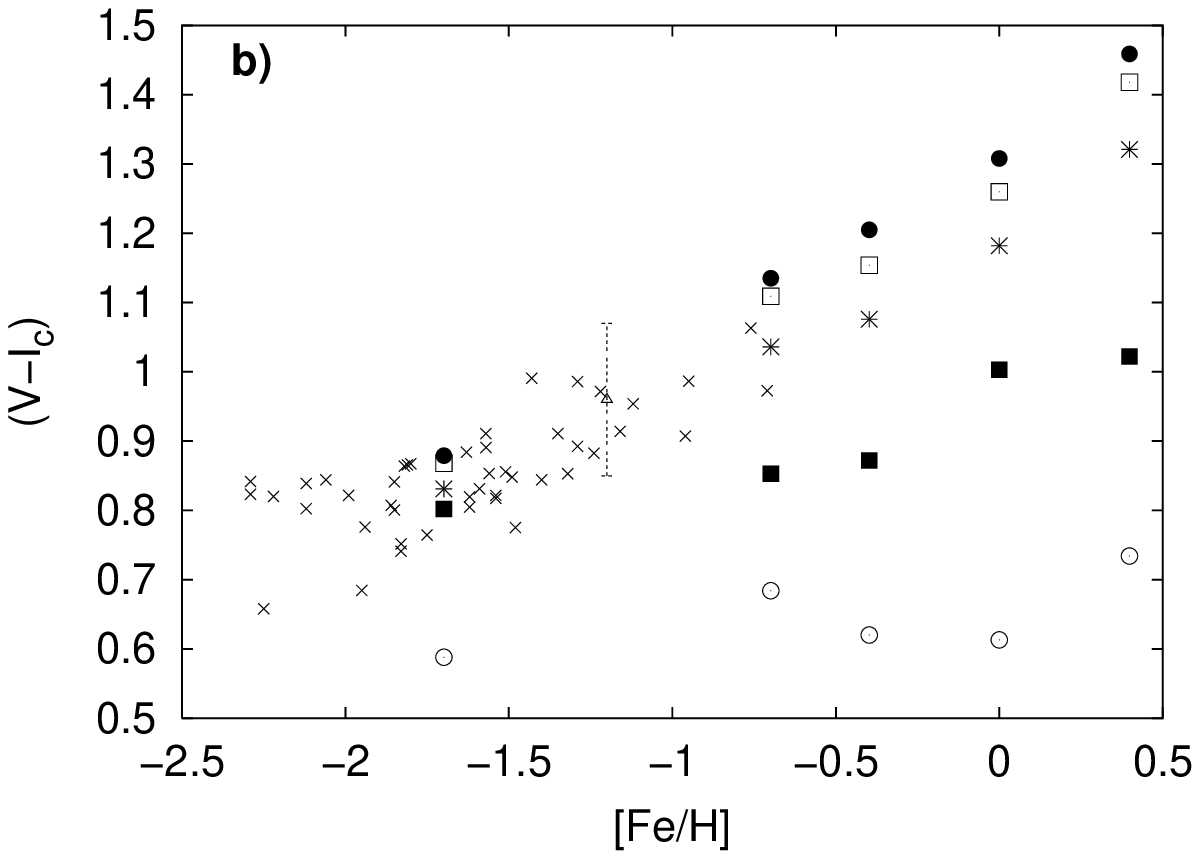}
\includegraphics[width=\columnwidth]{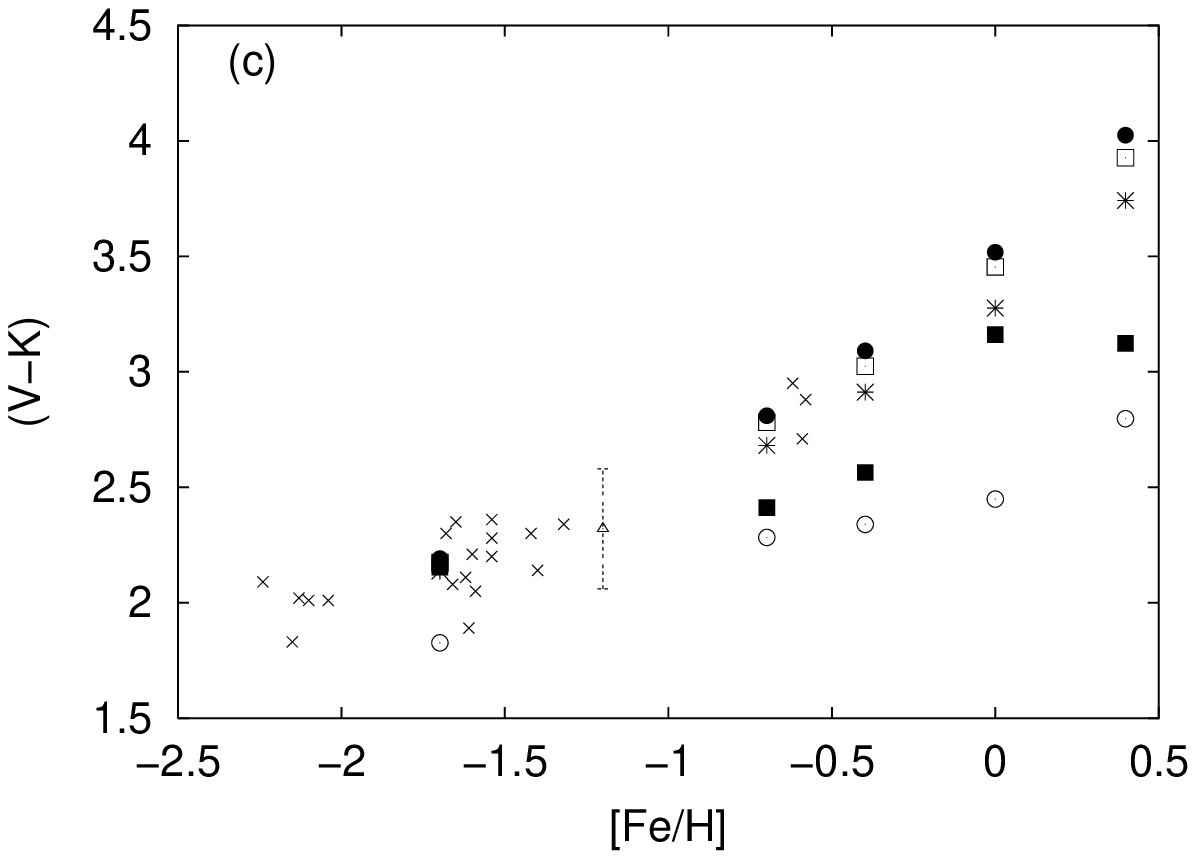}
\caption{{\bf a)} $(B-V)$, {\bf b)} $(V-I_{\rm c})$ ($I_{\rm c}$ is Cousins I)
and {\bf c)} $(V-K)$ color vs. metallicity [Fe/H] for SSPs with Salpeter
IMF at various ages. $0.5$ Gyr (open circles), $1$ Gyr (filled squares), $5$ Gyr (stars), $10$ Gyr (open squares), and $14$ Gyr (filled circles). 
Crosses in {\bf
a)} and {\bf b)} are Milky Way GC data from Harris' catalogue. In {\bf c)} the
crosses are M31 GC data from Brodie \& Huchra. Also included in all three figures are the mean M31 GC data and their $1\sigma$ ranges 
from Barmby \etal.} 
\label{z_sp}
\end{figure}

Clearly, some colors and color systems (like e.g. the Str\"omgren system)
are better suited for this purpose than others. It is important to note,
however, that in particular for optical and NIR colors the relations between color and metallicity are more or less age-dependent. This is seen in Figs
\ref{z_sp}a), b), where we plot theoretical calibrations obtained from our SSP
models at various ages for $(B-V)$ (Fig. \ref{z_sp}a)), 
$(V-I_{\rm c})$ (Fig. \ref{z_sp}b)) and $(V-K)$ (Fig. \ref{z_sp}c)) versus
[Fe/H]. We also include observations of GCs in the Milky way from the GC
catalogue compiled by W. E. Harris (\cite{harris}) (we only use clusters with
$E(B-V) < 0.4$) and median colors of GCs in M31 from Barmby \etal 
(\cite{barmby}) in Fig. \ref{z_sp} which show good agreement with our models at $10-14$ Gyr and low metallicities. In Fig. \ref{z_sp}c) observations from Brodie
\& Huchra (\cite{brodie}) are included which also show very good agreement with
our models. In this low metallicity regime empirical calibrations can be used
with good accuracy. In all three colors it is seen that at young ages the
theoretical calibrations are significantly different from what they are at old
ages. Strong changes are seen between 0.5 and 1 Gyr and between 1 and 5 Gyr, smaller changes at older ages, e.g. between 5 and 14 Gyr. Note that theoretical
calibrations can reach the highest metallicities for which stellar evolutionary
tracks or isochrones are available, i.e. $2.5~Z_{\odot}$ in our
case, while observational calibrations are restricted to the metallicity range
of Milky Way GCs, i.e. to $Z \lta 0.3~Z_{\odot}$.

Over the metallicity range of the GC sample $-2 \lesssim {\rm [Fe/H]}
\lesssim -0.5$ the agreement between our theoretical calibrations at 12 Gyr 
and the empirical ones from Couture \etal (\cite{cort}) is very good, i.e. better than $0.05$ mag both in $(B-V)$ and $(V-I_{\rm c})$.

Note, however, that extrapolations from the approximately linear observational
relation at [Fe/H] $\lta -0.5$ to higher metallicities would produce grossly
misleading results since towards higher metallicities the theoretical relations become significantly steeper.  By extrapolating the empirical linear relation,
a $10 - 14$ Gyr old cluster with e.g. $(B-V)=1.1$ would be ascribed
a metallicity $Z_{\rm extrap} \approx 4.4~Z_{\odot}$ while our model
shows it to only have $Z \approx 2.5~Z_{\odot}$. 

For $(V-I_{\rm c})$, the 
deviation from the linear relation valid in the low metallicity
region is even stronger. E.g. an old cluster with $(V-I_{\rm c})=1.2$ would
be ascribed $Z_{\rm extrap} \approx 3.0~Z_{\odot}$ while our models
indicate solar metallicity, i.e. three times lower! We stress that in particular
for investigations of intermediate age cluster populations or of clusters with
metallicities that might exceed those of the Milky Way or M31 GCs (on which the
empirical calibrations are based) it is very dangerous to use empirical
calibrations. Our models clearly show the age dependence as well as the strong
discrepancies from the empirical linear color -- metallicity relations for
[Fe/H] $\gta -0.5$.

\begin{figure}[ht]
\includegraphics[width=\columnwidth]{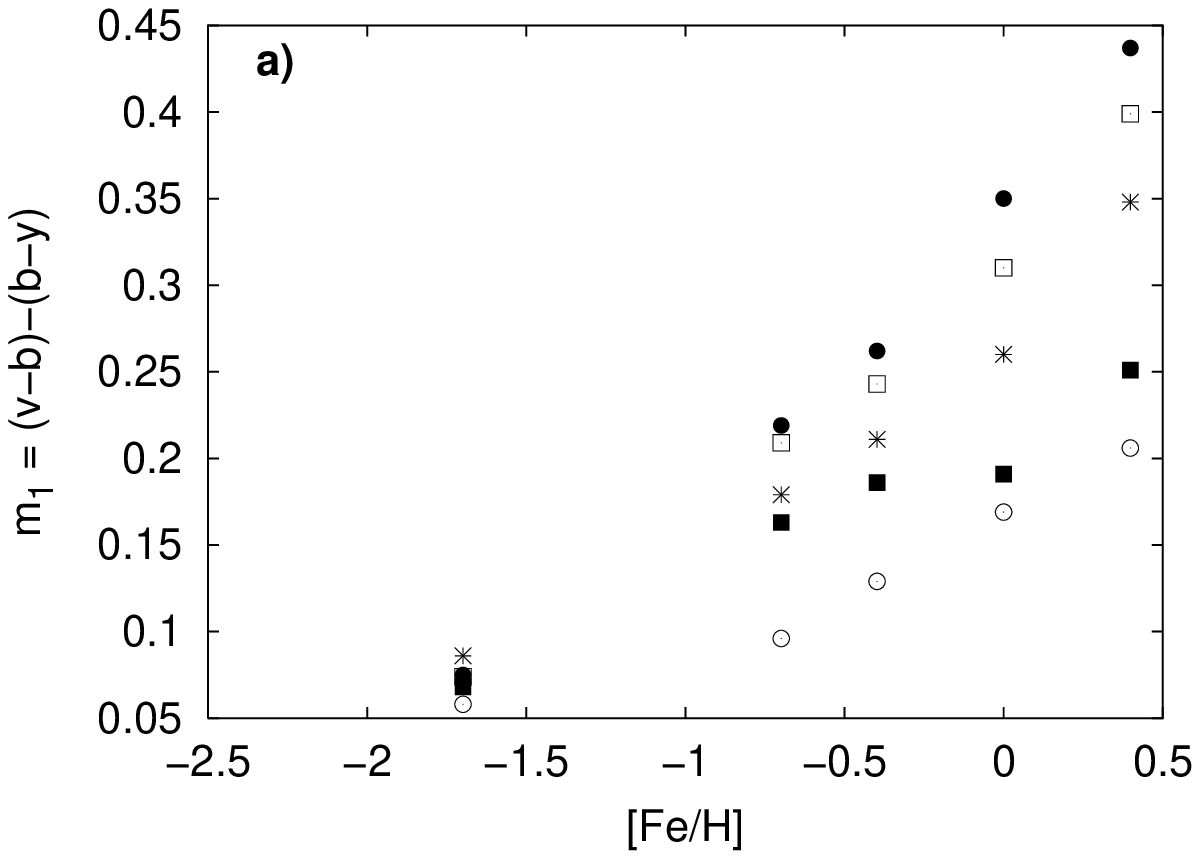}
\includegraphics[width=\columnwidth]{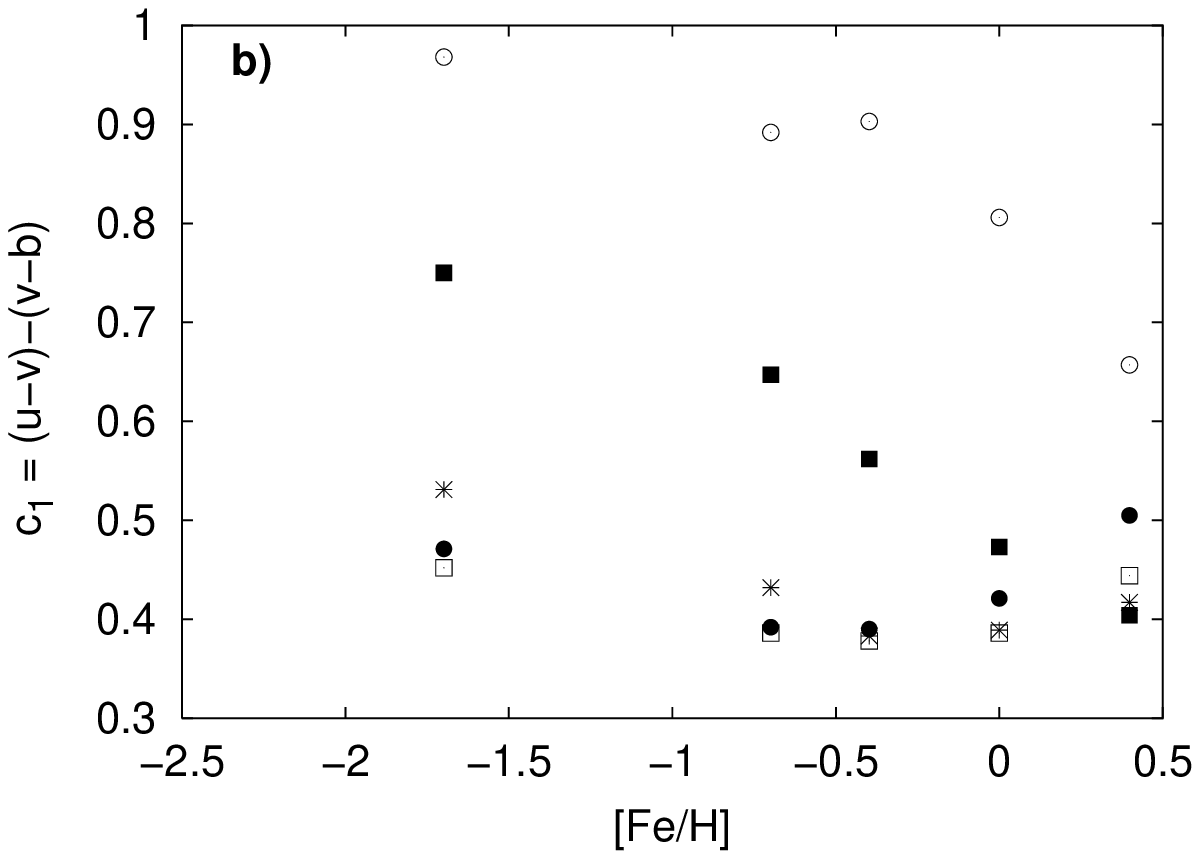}
\caption{{\bf a)} Str\"omgren $m_1$ and {\bf b)} $c_1$ color indices 
vs. metallicity [Fe/H] for SSPs with  Salpeter IMF at various ages. $0.5$ Gyr
(open circles), $1$ Gyr (filled squares), $5$ Gyr  (stars), $10$ Gyr (open
squares), and $14$ Gyr (filled circles).} 
\label{z_sp2}
\end{figure}

In Figs. \ref{z_sp2} a), b) we present analogues to Figs. \ref{z_sp} a), b),
c) for the Str\"omgren color indices $m_1= (v - b) - (b - y)$ and
$c_1 = (u - v) - (v - b)$. For the particularly metal-sensitive index
$m_1$ it is seen that it allows to well separate clusters in terms of
metallicities and ages. It allows for a better metallicity separation 
on the basis of colors than the Johnson
filter system does. But it has a smaller absolute difference, which makes it
fragile to observational uncertainties.

The age-sensitive Str\"omgren color index $c_1$ shows a completely
different behavior. While it is, indeed, relatively constant at given age for
all [Fe/H] $\lta -0.5$, it starts changing significantly towards higher metallicities, where it
allows for metallicity separations nearly as good as $m_1$. Due to its
changing behavior, it is important to well know the metallicity of the cluster,
which should be $< -0.5$ in [Fe/H]. Above this metallicity, even $m_1$
shows a better age separation than $c_1$.

While empirical calibrations like those obtained from Milky Way or M31 GC
analyses are only applicable for objects of comparatively old ages, our
theoretical calibrations can also be used for intermediate age and young
clusters, as long as emission lines from the active SF phase do no longer affect
the broad band colors. From our extensive electronic tables, theoretical
calibrations analogous to the ones depicted in Figs \ref{z_sp}a), b), c) can be
extracted for any color of our filter systems. They provide a useful tool for
the interpretation of young star clusters forming in large numbers in
interacting, merging, and starbursting gas-rich galaxies, for young and
intermediate age GCs in old merger remnants and kinematically young elliptical
and S0 galaxies. The extension to metallicities in the range $0.5~Z_{\odot}$ . . . $2.5~Z_{\odot}$, that is only
possible with theoretical calibrations, is of particular importance since
spectroscopy of the first handful of bright young star clusters recently formed
in spiral galaxy mergers has proven their relatively high metallicity --
compared to Galactic GCs (e.g. Schweizer \& Seitzer \cite{schwei}, Whitmore \etal \cite{w99}). Since these
$\mathrm{ 2^{nd}}$ generation GCs are formed from gas pre-enriched in the
interacting and/or starbursting galaxies, metallicities $\ge
0.5~ Z_{\odot}$ were expected (Fritze - v. A. \& Gerhard
\cite{fritze2}).

\section{Cosmological Evolution}
\label{sec:Cosm-Evol}
In this section, we present evolutionary and cosmological corrections for direct use of our SSP results in combination with galaxy formation or cosmological structure formation scenarios. We chose two
different cosmological models as characterised by $(H_0,~\Omega_0,~\Lambda_0)~=~ (65,~0.1,~0.0)$ and $(75,~0.1,~0.9)$,
respectively, and an assume an  onset of SF at redshift $z_{\rm f} = 5$. The manifold of cosmological models is constrained
by the ages of Milky Way GCs (12 -- 14 Gyr). 

In case of galaxies, the luminosities of model galaxies at today's age of 12 -- 14 Gyr are normalized to their observed average local
luminosities, i.e. to the average local $\langle M_B \rangle$ of the
respective galaxy type, and their evolutionary and cosmological corrections are
then given in terms of magnitude differences. Such a normalization is not
meaningful for SSPs if understood as galaxy constituents or
subpopulations. Hence, as explained in Sect.
\ref{sec:Cosmological-models}, we chose to present their evolutionary 
and cosmological corrections ($\epsilon_\lambda$, $\kappa_\lambda$) in 
terms of luminosity ratios in selected filter bands $\lambda$. 

In Table \ref{e-korr} we present these corrections as a function of redshift $z$ for the Johnson $U$ . . . $K$ bands. They are given for the two cosmological models chosen
above. Only the first few columns and lines of this table are shown for one cosmology in the printed version for orientation. Full tables are
available in the electronic version, at CDS, and our web site. 

The apparent magnitude $m_{\lambda}(z)$ in some filter $\lambda$ of an
SSP of given mass and metallicity at redshift $z$, $m_{\lambda}(z)~=~-2.5~
{\rm log}~L_{\lambda}(z,t(z))$ is obtained from the absolute luminosity $L_{\lambda}(0,t_0)$ of an SSP of given mass by today's age $t_0=t(z=0)$ via
\begin{eqnarray*}
m_{\lambda}(z)~=~-2.5~{\rm log}~(L_{\lambda}(0,t_0) \cdot \epsilon_{\lambda}(z) \cdot \kappa_{\lambda}(z)) + BDM(z).
\end{eqnarray*}

\begin{table}
\begin{tabular}[h]{cccccccc}
\hline
	$z$ & $\epsilon_U$ & $\kappa_U$ & $\epsilon_B$ & $\kappa_B$ &
$\epsilon_V$ &...\\
\hline 
0.03 & 1.0545 & 0.9175 & 1.0460 & 0.8815 & 1.0432 &\\
0.05 & 1.0870 & 0.8455 & 1.0726 & 0.7765 & 1.0658 &\\
0.08 & 1.1595 & 0.7653 & 1.1290 & 0.6863 & 1.1100 &\\
0.10 & 1.2110 & 0.6876 & 1.1570 & 0.6017 & 1.1200 &\\
0.13 & 1.2722 & 0.6121 & 1.1970 & 0.5275 & 1.1397 &\\
0.15 & 1.3439 & 0.5450 & 1.2510 & 0.4631 & 1.1738 &\\
0.17 & 1.4049 & 0.4767 & 1.2993 & 0.4103 & 1.2083 &\\
0.20 & 1.4646 & 0.4048 & 1.3455 & 0.3675 & 1.2442 &\\
0.23 & 1.5273 & 0.3400 & 1.3934 & 0.3296 & 1.2823 &\\
0.25 & 1.5904 & 0.2814 & 1.4434 & 0.2970 & 1.3226 &\\
\vdots \\
\hline
\end{tabular}
\caption{Evolutionary and cosmological corrections $\epsilon_{\lambda}$ and 
$\kappa_{\lambda}$, respectively, as a function of redshift $z$ from $z=0$ to $z=5$ in the
cosmology $H_0=65$, $\Omega_0=0.10$ for all Johnson bands for
an SSP with solar metallicity and Salpeter IMF. Full table with all wavelengths bands and for all metallicities and two different cosmologies is given in the electronic version, at CDS, and on our web page.}
\label{e-korr}
\end{table}

\section{Conclusions}
A new set of evolutionary synthesis models are presented for single burst single
metallicity stellar populations covering metallicities $0.02 \leq Z/Z_{\odot} 
\leq 2.5$ and ages $4\cdot 10^6 ~{\rm yr} \leq t \leq 16$ Gyr. They are
based on the most recent isochrones from the Padova group (Nov. 1999) that extend earlier
models by the inclusion of the thermal pulsing AGB phase for stars in the mass
range $2~ M_{\odot} \leq m \leq 7 ~M_{\odot}$ in accordance
with the fuel consumption theorem. We show that with respect to earlier models,
inclusion of the TP-AGB phase leads to significant changes in the $(V-I)$ and $(V-K)$ colors of SSPs in the age range from $10^8$ to $\gta 10^9$ yr, in particular at metallicities $Z \geq
0.5~Z_{\odot}$. The ages derived from observed $(V - I)$ colors of
young star clusters in this age range decrease by a factor of $\sim 2$ when the
TP-AGB phase is included. Using model atmosphere spectra from Lejeune \etal
(\cite{lej2}, \cite{lej}), we calculate the spectral evolution of single burst
populations of various metallicities covering the wavelength range from 90 {\AA}
~through 160 $\mu$m. Expanding dust shells around AGB stars, however, are not accounted for by our models. Therefore, they should not be used for the interpretation of IR data (i.e. beyond $K$) of intermediate age star clusters. Age and metallicity effects of the spectral and photometric
evolution of our SSP models are discussed in detail. Isochrone spectra are convolved with filter
response functions to describe the time evolution of luminosities and colors in
Johnson, Thuan \& Gunn, Koo, HST, Washington, and Str\"omgren filters.

Theoretical calibrations are provided for a number of colors in terms of
metallicity and compared with empirical calibrations and data from GCs. It is
found that the empirical calibrations are only suitable for the age and
metallicity range of the clusters they are derived from. Our models do not have
such restrictions and cover the whole metallicity -- and age -- range. 
They allow to study color -- metallicity calibrations in their time evolution 
and up to metallicities $\geq Z_{\odot}$, both of which is important 
for the interpretation of young and intermediate age star cluster systems. 

Model
results are not only intended for use in the interpretation of star clusters
over the full age range from few $10^7 - 1.4 \cdot 10^{10}$ yr but also for
combination with any kind of dynamical galaxy formation and/or evolution model
that contains a star formation criterion, i.e. some prescription how to transform gas into stars. Moreover, the evolution of these single burst
single metallicity stellar populations is readily folded with any kind of star formation --
and eventually chemical enrichment -- history to describe the evolutionary
spectral synthesis of composite stellar populations like galaxies of any type
with continuous or discontinuous SF.

For these latter purposes we also present the time evolution of mass-to-light
ratios (including ejection rates for gas and metals) for two different IMFs as
well as cosmological and evolutionary corrections for all the filters as a
function of redshift $z$ for $0 \leq z \leq 5$ and two different
cosmologies.

Extensive data files are provided in the electronic version, at CDS, and at http://www.uni-sw.gwdg.de/$\sim$galev/.

\begin{acknowledgements}
We thank Doug Geisler for his interest and help with the calibration of the
Washington system. 
\end{acknowledgements}

\end{document}